\documentclass[%
reprint,
 superscriptaddress,
nofootinbib,
 amsmath,amssymb,
 aps,
 prd,
floatfix,
]{revtex4-2}

\usepackage{graphicx,amssymb,amsmath,amsthm,amsfonts,epsfig,epsf}
\usepackage{url}
\usepackage[breaklinks]{hyperref}
\usepackage[usenames]{color}
\usepackage{epstopdf}
\usepackage{bm}
\usepackage{dcolumn}
\usepackage[latin1]{inputenc}
\usepackage{latexsym}
\usepackage{rotating}
\usepackage{color}
\usepackage{longtable}

\usepackage{enumerate}
\usepackage{tensor,multirow}
\usepackage{tikz,tikz-3dplot}
\usepackage[normalem]{ulem}

\usepackage{slashed}

\newcommand{\nn}{\nonumber}

\usepackage[scr=boondox]{mathalfa}

\DeclareFontFamily{U}{min}{}
\DeclareFontShape{U}{min}{m}{n}{<-> udmj30}{}

\newcommand{\milan}{\affiliation{Dipartimento di Fisica ``G. Occhialini'', 
Universit\`a degli Studi di Milano-Bicocca, Piazza della Scienza 3, 20126 Milano, Italy}}
\newcommand{\infn}{\affiliation{INFN, Sezione di Milano-Bicocca, 
Piazza della Scienza 3, 20126 Milano, Italy}}

\begin{document}


\title{Eccentric Catastrophes \& What To Do With Them}

\author{Nicholas Loutrel}
 \email{nicholas.loutrel@unimib.it}
 \affiliation{Dipartimento di Fisica, ``Sapienza'' Universit\`a di Roma \& Sezione INFN Roma1, Piazzale Aldo Moro 5, 00185, Roma, Italy}
 \milan
 \infn

\date{\today}

\begin{abstract}
Analytic modeling of gravitational waves from inspiraling eccentric binaries poses an interesting mathematical challenge. When constructing analytic waveforms in the frequency domain, one has to contend with the fact that the phase of the Fourier integral in non-monotonic, resulting in a breakdown of the standard stationary phase approximation. In this work, we study this breakdown within the context of catastrophe theory. We find that the stationary phase approximation holds in the context of eccentric Keplerian orbits when the Fourier frequency satisfies $f_{\rm min} < f < f_{\rm max}$, where $f_{\rm min/max}$ are integer multiples of the apocenter/pericenter frequencies, respectively. For values outside of this interval, the phase undergoes a fold catastrophe, giving rise to an Airy function approximation of the Fourier integral. Using these two different approximations, we generate a matched asymptotic expansion that approximates generic Fourier integrals of Keplerian motion for bound orbits across all frequency values. This asymptotic expansion is purely analytic and closed-form. We discuss several applications of this investigation and the resulting approximation, specifically: 1) the development and improvement of effective fly-by waveforms for binary black holes, 2) the transition from burst emission in the high eccentricity limit to wave-like emission in the quasi-circular limit, which results in an analogy between eccentric gravitational wave bursts and Bose-Einstein condensates, and 3) the calculation of f-mode amplitudes in eccentric binary neutron stars and black hole-neutron star binaries in terms of complex Hansen coefficients. The techniques and approximations developed herein are generic, and will be useful for future studies of gravitational waves from eccentric binaries within the context of post-Newtonian theory.
\end{abstract}

\maketitle



\section{Introduction}
\label{sec:intro}

The emission of gravitational waves (GWs) from binary systems comprised of compact objects generally causes the orbital eccentricity to decay~\cite{Peters:1963ux,Peters:1964zz}. While the isolated binary black hole (BBH) formation channel, consisting of two mutually evolving giant stars, will typically lead to binaries with negligible eccentricity in the detection band of ground-based GW detectors~\cite{Mapelli:2020:review,Sedda:2020:fingerprints,Belczynski:2015tba,Belczynski01,Grudzinska:2015eta,Kowalska:2010qg}, dynamically assembled BBHs formed in dense stellar environments and AGN disks can possess arbitrarily large orbital eccentricity~\cite{Tagawa:2020jnc,Tagawa:2020qll,SamsigRamirezRuiz:2017:HighlyEccentric,Samsing:2014:BinarySingle,Samsing:2017xmd,Samsing:2018isx,Samsing:2018ykz,Samsing:2020:AGN,Samsing:2020:massgap,Gayathri:2021xwb}. Eccentricity thus provides a clean indicator of the origin of GW signals from compact binary coalescences (CBCs), baring the possibility of confusion with other relativistic two-body effects, such as precession~\cite{Romero-Shaw:2022fbf}, and detector noise~\cite{Xu:2022spd}. Indeed, reanalysis of the signals already detected have found signatures of eccentricity in four of the confirmed detections, providing the first evidence that some of the sources already observed are dynamically assembled~\cite{Romero-Shaw:2019:GWTC-1-ecc,Romero-Shaw:2021:GWTC-2-ecc}.

While the development of waveform templates for eccentric binaries has historically lagged behind other binary sources, the topic is receiving steadily increasing attention. Analytic models of the inspiral phase have been developed for binaries with eccentricity $e \lesssim 0.6$ to third post-Newtonian (PN) order~\cite{Moore:2018kvz,Moore:2019xkm}, and in the high eccentricity ($e\sim1$) limit at leading PN order (so-called Newtonian order in both conservative and dissipative dynamics)~\cite{Loutrel:2019kky,Gallouin:2012kb}. Significant progress has also been made to extend the effective one-body (EOB) waveforms to arbitrary eccentricity~\cite{Nagar:2021:EccentricWaveform,RamosBuades:2022:EccentricWaveform,Liu:2022:EccentricWaveform}, as well as extend the PN inspiral-only waveforms to full inspiral-merger-ringdown (IMR) waveforms~\cite{Cho:2022:EccentricWaveform}. However, presently it is difficult to quantify exactly how accurate these waveform models are in the high eccentricity regime due to the lack of an ``exact" waveform, specifically those provided by numerical relativity (NR). While initial data of relevance to BBHs generally results in a non-negligible amount of orbital eccentricity, computational limitations have historically prevented the simulation of binaries with moderate and high eccentricity beyond a few orbital cycles~\cite{Stephens:2011as,East:2011xa,East:2012ww}. Recent progress toward addressing this may be found in~\cite{Boyle:2019kee}.

Within the context of PN theory, the analytic modeling of eccentric binaries has proven to be mathematically rich, and challenging. Due to Fourier space reducing the complexity of GW data analysis, the end goal of analytic waveform modeling is usually to develop a closed-form expression (i.e. one that does not require an infinite summation of terms, or evaluation of numerical integrals) for the waveform template in terms of frequency rather than time. Even in the time domain, modeling of eccentric orbits can be challenging, and it is well known even at the Newtonian level that the dynamics can generically only be reduced to quadratures~\cite{PoissonWill}. Explicit solutions require either perturbative techniques~\cite{Yunes:2009yz} or Fourier series methods~\cite{Moore:2018kvz,Moreno-Garrido}.

When transforming to the frequency domain, one has to solve an integral of an oscillatory function with complicated phase behavior~\cite{Moore:2018kvz}. The general technique for evaluating such integrals is the application of the stationary phase approximation (SPA)~\cite{Bender}. However, for eccentric binaries, the SPA generally breaks down, unless one performs a suitable transformation of the integrand to avoid singularities in the approximation, as was done in~\cite{Moore:2018kvz,Moore:2016qxz,Klein:2014bua}. The same procedure was used in~\cite{Loutrel:2019kky,Loutrel:2020jfx} to develop effective fly-by (EFB) waveforms, which aim to model the GW bursts from highly eccentric binaries. While these methods have been very useful for developing waveform models of low and moderately eccentric binaries to high PN order, they have failed for highly eccentric binaries due to the complicated structure of the PN-extended Fourier series description of the two-body problem~\cite{Boetzel:2017zza}. As a result, the high eccentricity EFB waveforms of~\cite{Loutrel:2019kky,Loutrel:2020jfx} have not been extended beyond leading PN order.

In this paper, we take a step toward resolving this problem, while also simultaneously elucidating the complicated phase and frequency structure of GWs from eccentric binaries. The Fourier integrals in question possess a phase function of the form $\Psi = 2\pi f t - m V$, with time $t$, frequency $f$, true anomaly of the orbit $V$, and harmonic number $m$, which only takes integer values. This phase function mimics those found in the study of the full inspiral of eccentric CBCs. We show that stationary points only exists when a particular condition is satisfied, specifically when the Fourier frequency is between the apocenter and pericenter frequencies, multiplied by the integer $m$. When the Fourier frequency equals either of these, the SPA obtains a singularity and becomes divergent. Such a singularity is referred to as a \textit{catastrophe}, and the study of such quantities is known as \textit{catastrophe theory}~\cite{Thom,ROOPNARINE2008531,Arnold,Golubitsky}. In the high frequency limit, specifically when the Fourier frequency is greater than the pericenter frequency, the SPA is no longer valid, and the Fourier integrals are approximated by an Airy function response. We develop a matched asymptotic expansion across the catastrophe, which provides a closed-form and analytic expression for generic Fourier integrals of Keplerian quantities.

After developing the matched asymptotic expansion, we investigate a number of applications of this result. First, we consider the construction of new EFB waveforms. By computing the match (or faithfulness)~\cite{Buonanno:2009zt} between numerical PN waveforms and the new EFB waveforms, we find that the analytic waveforms are a faithful representation of waveforms that may exist in nature. Second, we use the investigation of the critical points of the phase function $\Psi$ to characterize the behavior of GWs from inspiraling eccentric binaries, which presents an intriguing connection with Bose-Einstein condensates. Lastly, we show that the asymptotic expansions can be used to approximate Hansen coefficients, which are quantities appearing in the Fourier series description of generic Keplerian orbital quantities~\cite{Hansen}.

The remainder of the paper is organized as follows. In Sec.~\ref{sec:cat}, we provide a brief overview of catastrophe theory to provide an introduction to the methodology used in following sections. In Sec.~\ref{sec:ecc}, we define the Fourier integral $E_{m}^{\pm}(f)$ under investigation, and study the SPA and high frequency limit in Secs.~\ref{sec:spa} \&~\ref{sec:fold}, respectively. We develop the matched asymptotic expansion in Sec.~\ref{sec:matched}, and discuss its applications in Sec.~\ref{sec:apps}. Finally, we discuss future directions in Sec.~\ref{sec:disc}. Throughout this work, we use units where $G = c = 1.$

\section{Catastrophe Theory: A Primer}
\label{sec:cat}

Consider a generalized Fourier integral of the form
\begin{equation}
    \label{eq:int}
    I(x) = \int_{a}^{b} A(t) e^{i x \Psi(t)} dt\,,
\end{equation}
where $[A(t), \Psi(t)]$ are smooth arbitrary functions of $t$. Such integrals appear frequently in physical applications, a few being general diffraction problems~\cite{BerryNye,M_V_Berry_1977,BerryUmbilic,BERRY1980257}, radio astronomy~\cite{Cordes:2017eug,Main:2018kfc,Feldbrugge:2019fjs,Melrose_2006}, and quantum path integrals~\cite{Tanizaki:2014xba,Behtash:2015loa}. For some choices of the functions $[A(t), \Psi(t)]$ the integral is known explicitly in closed form. However, in general, this is not true and one typically has to look for approximate solutions. The methods of obtaining such solutions fall into the purview of asymptotic analysis. The most common method employed is that of the \textit{stationary phase approximation} (SPA)~\cite{Bender}, wherein one searches for stationary points of the phase defined as
\begin{equation}
    \dot{\Psi}(t_{\star}) = 0
\end{equation}
where the over dot corresponds to differentiation with respect to time $t$. The time $t$ that solves this is the stationary point $t_{\star}$. When the stationary point exists, the integral in Eq.~\eqref{eq:int} becomes dominated by the region around $t_{\star}$ and it is suitable to Taylor expand both the phase and amplitude, specifically
\begin{align}
    \Psi(t) &\sim \Psi(t_{\star}) + \frac{1}{2} \ddot{\Psi}(t_{\star}) (t-t_{\star})^{2} + {\cal{O}}\left[(t-t_{\star})^{3}\right]\,,
    \\
    A(t) &\sim A(t_{\star}) + {\cal{O}}\left(t-t_{\star}\right)\,.
\end{align}
The integral can now be evaluated by taking the limits of integration to infinity, which is acceptable since the integrand oscillates rapidly outide of the region around the stationary point $t_{\star}$, and thus will evaluate to a small number. The end result is
\begin{equation}
    \label{eq:spa}
    I(x) \sim \sqrt{\frac{2\pi}{x |\ddot{\Psi}(t_{\star})|}} A(t_{\star}) e^{i x \left[ \Psi(t_{\star}) + {\rm sign}[\ddot{\Psi}(t_{\star})] \pi/4\right]}
\end{equation}
which is the simplest version of the SPA.

Now, suppose that the phase function in not simply a function of time $t$, but also of a parameter $\lambda$, i.e. $\Psi = \Psi(t,\lambda)$. We can still search for stationary points satisfying
\begin{equation}
    \dot{\Psi}(t_{\star},\lambda) = 0\,,
\end{equation}
but now the stationary point will be manifestly a function of the parameter, specifically $t_{\star} = t_{\star}(\lambda)$. Typically, this is not an issue, and one simply has to promote all functions of $t_{\star}$ in Eq.~\eqref{eq:spa} to functions of $\lambda$. However, if there is a value of $\lambda = \lambda_{c}$ such that
\begin{equation}
    \ddot{\Psi}(\lambda_{c}) = \ddot{\Psi}[t_{\star}(\lambda_{c}),\lambda_{c}] = 0\,,
\end{equation}
then the SPA given by Eq.~\eqref{eq:spa} diverges at $\lambda_{c}$, forming a catastrophe. 

\begin{figure*}[hbt!]
    \includegraphics[width=\textwidth, trim={4cm 0cm 4cm 2cm}, clip]{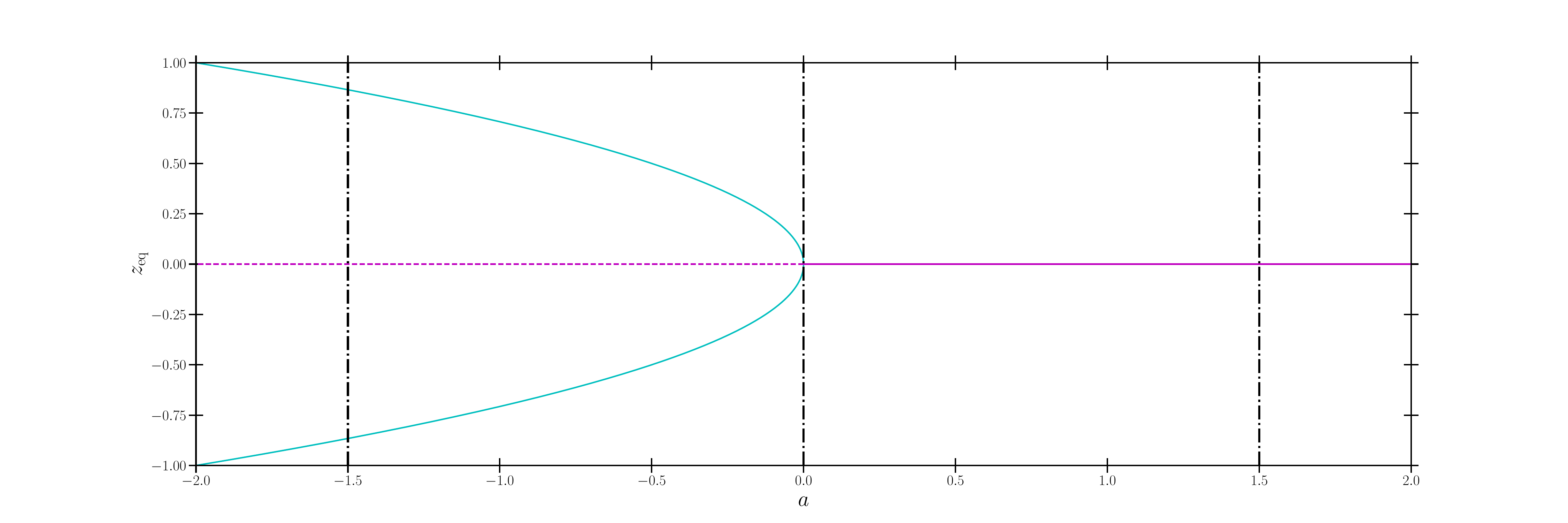}
    \includegraphics[width=\textwidth, trim={4cm 0cm 4cm 2cm}, clip]{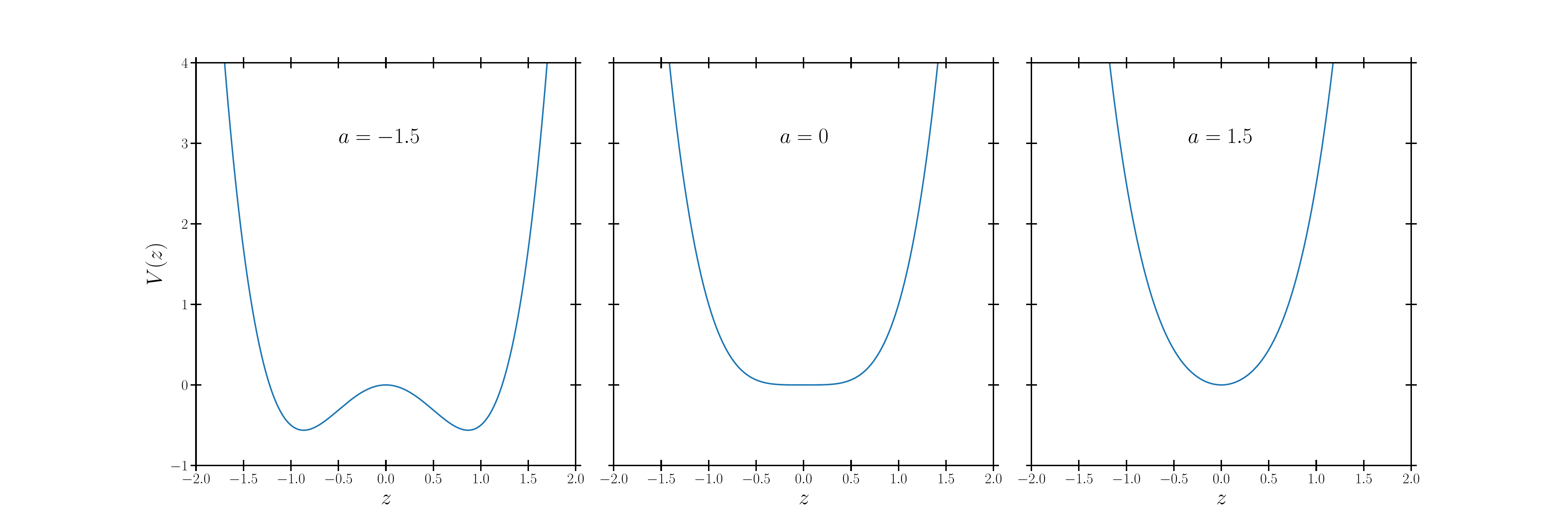}
    \caption{\label{fig:part} Top: Bifurcation diagram showing the location of equilibria for a classical particle moving in a one-dimensional potention $V(z,a) = z^{4} + a z^{2}$, with parameter $a$. The style of each line indicates whether the equilibria are stable (solid) or unstable (dashed). For $a>0$, only one equilibrium exists at $z=0$ (magenta line). For $a<0$, two stable equilibria exist at $z = \pm \sqrt{|a|/2}$ (cyan line), while the equilibrium at $z=0$ becomes unstable. The transition point at $a=0$ corresponds to a fold catastrophe. The vertical dot-dashed lines show select values of $a$ that are used in the bottom plots. Bottom: Plot of the potential $V(z,a)$ for different values of $a$ corresponding to the vertical dot-dashed lines in the bifurcation diagram. From left to right, $a=[-1.5,0,1.5]$.}
\end{figure*}

The types of catastrophes relevant to the topic of this paper are known as \textit{fold catastrophes}. A simple example that elucidates this phenomenon is a classical particle moving in a one dimensional ``sombrero" potential
\begin{equation}
    V(z,a) = z^{4} + a z^{2}\,,
\end{equation}
where $a$ is a real-valued parameter, and $z$ is the position of the particle. The equilibria of this system are found by solving
\begin{equation}
    V'(z,a) = 4 z^{3} + 2 a z = 0\,,
\end{equation}
where the prime correponds to differentiation with respect to $z$. The nature of the equilibria (which are stationary points), and as a result, the dynamics of the particle, depend on the value of the parameter $a$. When $a$ is positive, there is only one stationary point, namely $z=0$, which is stable since $V''(0,a) > 0$. As $a$ decreases and reaches zero, the potential becomes increasingly flat at the stationary point. However, when $a$ is negative, the behavior of the equilibrium changes. The necessary equation to solve for the equilibria is now
\begin{equation}
    4 z^{3} - 2 |a| z = 0\,,
\end{equation}
for which there are now three stationary points, specifically $z=0$ and $z = \pm \sqrt{|a|/2}$. The two new equilibria are stable, while the previous equilibrium at $z=0$ is now unstable, and the particle will ``decay" to one of the stable equilibria under small perturbations. This sudden change in behavior of the system is the reason why the point $a=0$ is called a catastrophe. Fig.~\ref{fig:part} provides a visual graphic of this behavior, along with a bifurcation diagram in the bottom panel showing how the equilibria evolve as a function of the parameter $a$. 

The above example of a fold catastrophe is a basic model of spontaneous symmetry breaking~\cite{Beekman:2019pmi}, a phenomena that appears frequently in many branches of physics, not least of which is the Higgs mechanism in particle physics. The behavior of many dynamical systems can be well understood within the context of catastrophe theory, and the discussion presented in this section is merely a simple introduction to the topic for the purpose of providing background on the problem at hand. However, before moving on, it is important to note that catastrophe theory does not provide a tool set of how to deal with catastrophes in the setting of Fourier transforms. In fact, catastrophe theory merely provides a means of classifying the singular points of a dynamical system, with the end goal of obtaining a deeper understanding of the behavior of the system. In order to properly approximate the behavior of a dynamical system through a catastrophe in a uniform manner, the tools provided by \textit{asymptotic analysis} are typically required. Since asymptotic analysis is a broad topic, we simply point the reader to the following text on the topic~\cite{Bender}.

\section{Eccentric Catastrophes}
\label{sec:ecc}

In this section, we discuss the presence of catastrophes in the context of eccentric binaries on Keplerian orbits. While the discussion is limited to Newtonian (or leading PN) order, it forms the basis necessary to analyze binary dynamics in general relativity (GR) within the context of PN theory. 

\subsection{Keplerian Orbits \& Definitions}
\label{sec:fourier}

Keplerian orbits describe the motion of two bodies orbiting around a common center of mass within Newtonian gravity~\cite{PoissonWill}. The motion of two point masses in Newtonian gravity has sufficient symmetries that the motion can be confined to a plane, spanned by the coordinates $(r,\phi)$ and with a normal described by the orbital angular momentum vector $\vec{L}$. A sufficient solution in the form of quadratures for the motion is given by
\begin{align}
    \label{eq:r12}
    r_{12} &= \frac{p}{1 + e \cos V}\,,
    \\
    \label{eq:Vdot}
    \dot{V} &= \Omega \left(1 + e \cos V\right)^{2}\,,
\end{align}
where $r_{12}$ is the radial separation of the two objects, $V = \phi_{12} - \omega$ is the true anomaly with $\phi_{12}$ the orbital phase and $\omega$ the longitude of pericenter, $\Omega = (M/p^{3})^{1/2}$ with $M$ the total mass of the binary, and $(p, e)$ are the semi-latus rectum and eccentricity of the orbit respectively. In the absence of perturbations, $(p,e)$ are constants of motion and are directly related to the orbital energy $E$ and magnitude of the orbital angular momentum $L = |\vec{L}|$ by
\begin{equation}
    E = -\frac{\mu M}{2 p}(1-e^{2})\,, \qquad L = \mu \left(M p\right)^{1/2}\,,
\end{equation}
where $\mu=m_{1}m_{2}/M$ is the reduced mass of the binary, with $m_{1,2}$ the binary component masses.

Now, consider the following Fourier integral,
\begin{align}
    \label{eq:Em-int}
    E_{m}^{\pm}(f) = \int_{-\pi}^{\pi} \frac{d\ell}{n} e^{\pm i m V} e^{2\pi i f t}\,,
\end{align}
and it's associated integrals,
\begin{align}
    C_{m}(f) &= \int_{-\pi}^{\pi} \frac{d\ell}{n} \cos(m V) e^{2\pi i f t} = \frac{1}{2} \left[ E^{+}_{m}(f) + E^{-}_{m}(f)\right]\,,
    \\
    S_{m}(f) &= \int_{-\pi}^{\pi} \frac{d\ell}{n} \sin(m V) e^{2\pi i f t} = \frac{1}{2i} \left[ E^{+}_{m}(f) - E^{-}_{m}(f) \right]\,.
\end{align}
In the above expressions, $\ell = n (t-t_{p})$ in the mean anomaly, and we take $m>0$. These integrals can be thought of as short-time Fourier transforms, where the time interval is given by a single orbital cycle, rather than all time. Such quantities appear frequently in the study of eccentric binaries within the PN formalism~\cite{Moreno-Garrido,Moore:2018kvz}, and are actually a limit of the more general Hansen coefficients~\cite{Hansen} which will be made clearer in Sec.~\ref{sec:hans}. Due to the complicated nature of the true anomaly $V(t)$, typically there are no closed-form expressions for these integrals except for some exceptional values of $m$. Thus, we must resort to approximate methods of solving the integral in Eq.~\eqref{eq:Em-int}.

\subsubsection{Stationary Phase Approximation}
\label{sec:spa}

The integral in Eq.~\eqref{eq:Em-int} can be directly mapped to the generalized Fourier integral in Eq.~\eqref{eq:int}. As such, the same techniques for evaluating it apply, and we begin by searching for any stationary points of the phase defined by $\Psi_{\pm}(t) = 2\pi f t \pm m V(t)$, with $V(t)$ given by Eq.~\eqref{eq:Vdot}. The stationary points will satisfy $\dot{\Psi}_{\pm} = 0$, which gives the equation
\begin{equation}
    2\pi f \pm m \Omega \left(1 + e \cos V_{\star}\right)^{2} = 0\,,
\end{equation}
with $V_{\star} = V(t_{\star})$ being the stationary point. The two solutions for $V_{\star}$ are then
\begin{align}
    \label{eq:Vsp}
    V_{\star}^{(1)} = \cos^{-1}\left[\frac{1}{e} \left(\sqrt{\frac{\mp 2\pi f}{m\Omega}} - 1\right)\right]\,, \qquad V_{\star}^{(2)} = - V_{\star}^{(1)}\,.
\end{align}
The existence of stationary points depends on the values of $(f,e)$ for any given $\Omega$. First, the stationary points only exist for positive frequencies for the $\Psi_{-}$ (the `$+$' sign in Eq.~\eqref{eq:Vsp}) and for negative frequencies for $\Psi_{+}$ (the `$-$' sign in Eq.~\eqref{eq:Vsp}). For the remainder of the discussion, we will focus on $\Psi_{-}$, since the calculation for $\Psi_{+}$ follows the same steps, but only for negative frequencies. We provide a suitable approximation for $E_{m}^{+}(f)$ for positive frequencies in Appendix~\ref{app:Em_plus}.

The above considerations indicate that the argument of the inverse cosine in Eq.~\eqref{eq:Vsp} must be between $[-1,1]$. For any given $(\Omega,e)$, this means that the frequency $f$ must be between $f_{\rm min}$ and $f_{\rm max}$ in order for the stationary points to exist, where
\begin{align}
    \label{eq:fmin-max}
    f_{\rm min} &= \frac{m\Omega}{2\pi} (1-e)^{2}\,, \qquad f_{\rm max} = \frac{m\Omega}{2\pi} (1+e)^{2}\,.
\end{align}
Note that these are integer multiples of the Fourier frequencies of apastron and periastron, respectively. When $f=f_{\rm min}$, the stationary points sit at the edge of the domain of integration, specifically $V_{\star}^{(1)} = - V_{\star}^{(2)} = \pi$. When $f=f_{\rm max}$, the stationary points coalescence and become $V_{\star}^{(1)} = V_{\star}^{(2)} = 0$. For $f_{\rm min} < f < f_{\rm max}$, the SPA is a valid approximation and Eq.~\eqref{eq:Em-int} evaluates to
\begin{equation}
    \label{eq:Em-spa}
    \left[E_{m}^{-}(f)\right]_{\rm SPA} = 2 \sqrt{\frac{2\pi}{\ddot{\Psi}_{-,\star}}} \cos\left(\frac{\pi}{4} - \Psi_{-,\star}\right) \,,
\end{equation}
where
\begin{align}
    \label{eq:spa-phase}
    \Psi_{-,\star} &= 2\pi f t\left(V_{\star}^{(2)}\right) - m V_{\star}^{(2)}\,,
    \\
    \label{eq:spa-amp}
    \ddot{\Psi}_{-,\star} &= 4 e \Omega^{2} \left(\frac{2\pi f}{m \Omega}\right)^{3/2} \left[1 - \frac{1}{e^{2}} \left(\sqrt{\frac{2\pi f}{m \Omega}} - 1\right)^{2} \right]^{1/2}
\end{align}
and we have summed over the contribution from both stationary points to obtain Eq.~\eqref{eq:Em-spa}. This constitutes the SPA of Eq.~\eqref{eq:Em-int}.

\begin{figure*}[hbt]
    \centering
    \includegraphics[width=\textwidth, trim={4cm 0cm 4cm 2cm}, clip]{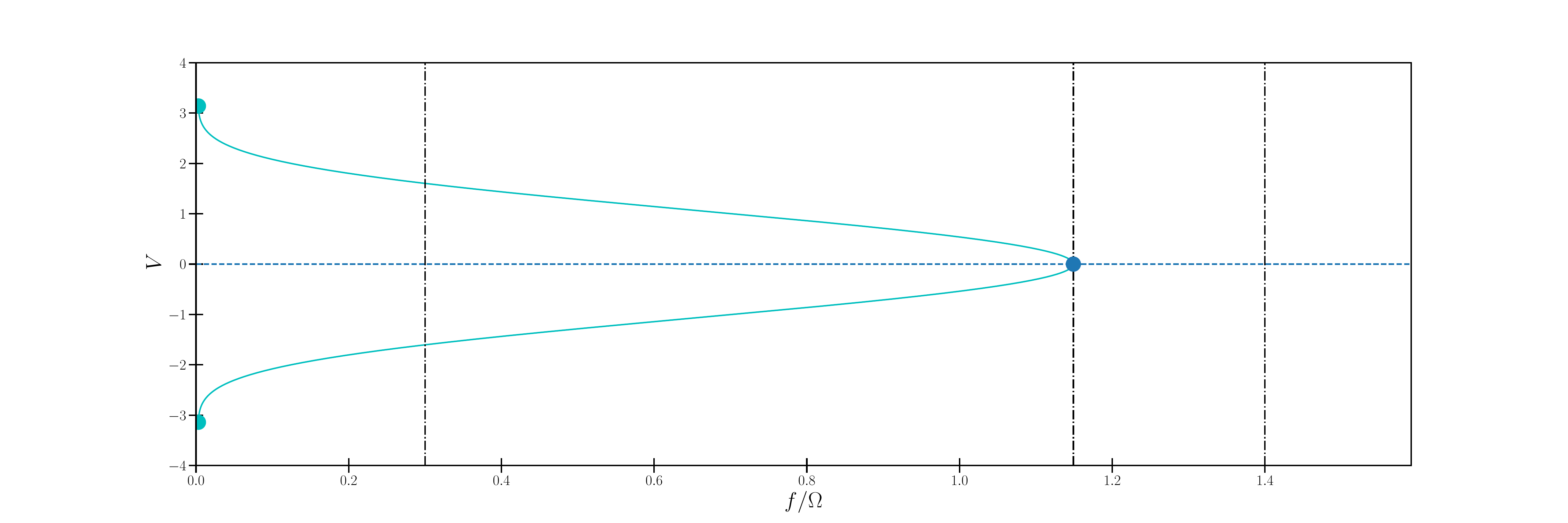}
    \includegraphics[width=\textwidth, trim={4cm 0cm 4cm 2cm}, clip]{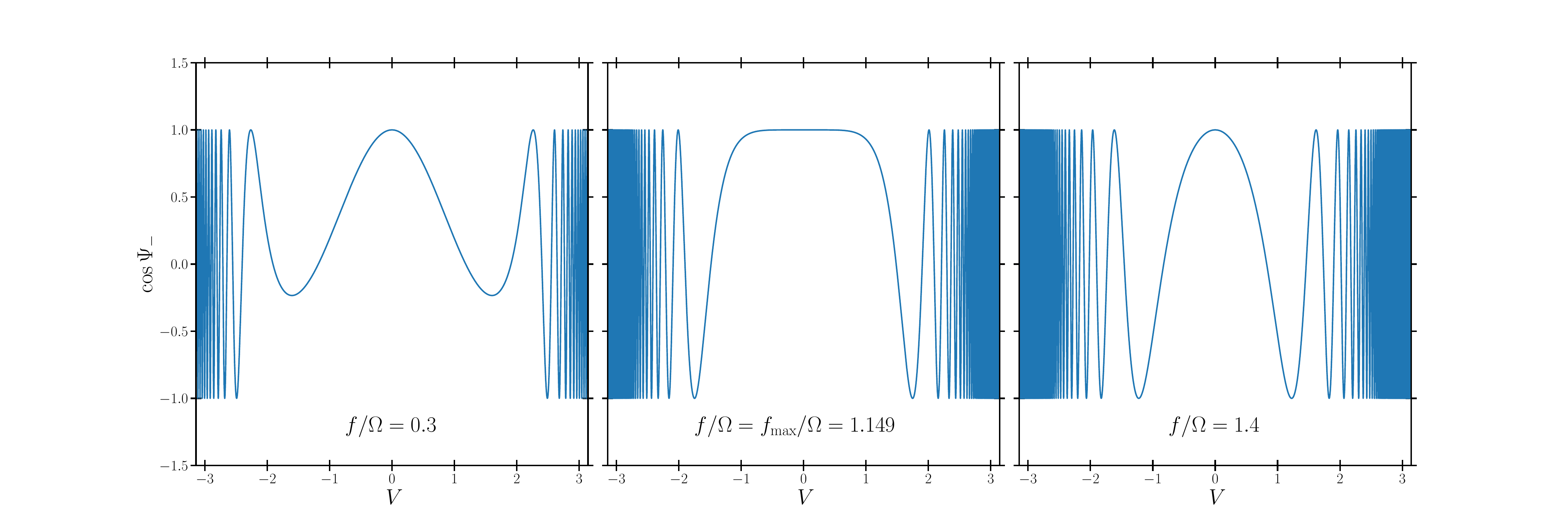}
    \caption{\label{fig:Em-plot} Top: Bifurcation diagram of the phase $\Psi_{-}(t) = 2\pi f t - m V(t)$ for $m=2$ and $e=0.9$ as a function on $f/\Omega$. The stationary points (cyan lines) appear at $V = \pm \pi$ (cyan points) at $f_{\rm min}$. As the frequency increases, the stationary points move toward the origin, and finally coalesce at $V=0$ when $f=f_{\rm max}$ (blue point). At higher frequencies, the stationary points disappear, and the saddle point (blue dashed line) dominates the phase. The vertical dot-dashed lines show particular values of $f/\Omega$ that are displayed in the bottom plots. 
    Bottom: Plot of $\cos\Psi_{-}$ for $m=2$ and at select values of $f/\Omega$ corresponding to the black dot-dashed lines in the top plot, specifically $f/\Omega = [0.3, 1.149, 1.4]$ from left to right.}
\end{figure*}

It is straightforward to show that $\ddot{\Psi}_{-,\star}(f=f_{\rm min/max}) = 0$, and thus the SPA given in Eq.~\eqref{eq:Em-spa} possesses catastrophes at $f=f_{\rm max}$ and $f=f_{\rm min}$. Because $f_{\rm min}$ is the frequency associated with apastron, it is typically very small, especially in the context of highly eccentric GW burst sources for ground based detectors. Here, we will primarily concern ourselves with the catastrophe that occurs at $f=f_{\rm max}$. It is worth pointing out that there is a point where $\dot{\Psi}_{-} \neq 0$, but $\ddot{\Psi}_{-} = 0$, specifically the saddle point $V = V_{c} = 0$. When $f=f_{\rm max}$, the stationary points coalescence with $V_{c}$, creating the relevant fold catastrophe. At higher frequencies, the stationary points disappear, and the integral of $E_{m}^{-}$ is dominated by the saddle point. The top panel of Fig.~\ref{fig:Em-plot} provides a bifurcation diagram that explicitly shows the evolution of the stationary points (solid line) for an example binary with $e=0.9$ and harmonic number $m=2$. The catastrophe is displayed by the blue circle at $f/\Omega = 1.149$, with $V_{c}$ shown in the dashed line. Note that $V_{c}$ is present at all frequencies, but it's contribution to $E_{m}^{-}$ is subdominant below $f_{\rm max}$. 

\subsubsection{High Frequency Approximation}
\label{sec:fold}

Since it is clear now that the SPA fails when $f \ge f_{\rm max}$, how does one approximate the behavior of the integral in Eq.~\eqref{eq:Em-int}? We follow a similar procedure to the SPA, but we instead expand about the saddle point at $V=0=\ell$, which corresponds to $t=t_{p}$ (i.e. periastron). The expansion of $V$ near periastron can be achieved using repeated differentiation of Eq.~\eqref{eq:Vdot}, specifically
\begin{equation}
    V(\psi) = (1+e)^{2} \psi - e (1+e)^{5} \frac{\psi^{3}}{3} + {\cal{O}}(\psi^{5})\,,
\end{equation}
where $\psi = \ell/\epsilon^{3/2} = \Omega (t-t_{p})$. Applying this to Eq.~\eqref{eq:Em-int}, we have
\begin{equation}
    \label{eq:Em-fold}
    E_{m}^{-}(f) \sim 
    \frac{e^{2\pi i f t_{p}}}{\Omega} \int_{-\infty}^{\infty} d\psi \; e^{i [\Delta(f) \psi + \sigma \psi^{3}/3]}
\end{equation}
with
\begin{equation}
    \label{eq:sigma-def}
    \Delta(f) = \frac{2\pi f}{\Omega} - m (1+e)^{2}\,, \qquad \sigma = m e (1+e)^{5}\,.
\end{equation}
Note that $\Delta(f) \ge 0$ due to the fact that we are working in the limit $f \ge f_{\rm max}$. Further, we have taken the limits of integration to infinity, in the same manner as the SPA. The integral in Eq.~\eqref{eq:Em-fold} can be mapped to the integral definition of the Airy function ${\rm Ai}(x)$~\cite{Abromowitz,NIST:DLMF}, and thus
\begin{equation}
    \label{eq:Em-airy}
    \left[E_{m}^{-}(f)\right]_{\rm high-f} = \frac{2\pi}{\sigma^{1/3}\Omega} e^{2\pi i f t_{p}} {\rm Ai} \left[\Delta(f)/\sigma^{1/3}\right]\,,
\end{equation}
which provides a sufficient approximation for $f\gg f_{\rm max}$.

Before proceeding, it is worth understanding the asymptotic behavior of the above expression. For large arguments, the Airy function exponentially decays. Thus, when $f\gg f_{\rm max}$, Eq.~\eqref{eq:Em-airy} becomes
\begin{equation}
    E_{m}^{-}(f) \sim \frac{e^{-(2/3)z^{3/2}}}{z^{1/4}}\,,
\end{equation}
where $z = \Delta(f)/\sigma^{1/3}$. While the methods to obtain Eq.~\eqref{eq:Em-airy} assumed $f\ge f_{\rm max}$, Eq.~\eqref{eq:Em-airy} is regular for $f<f_{\rm max}$, and it is instructive to take this limit as well. When $f\ll f_{\rm max}$, Eq.~\eqref{eq:Em-airy} becomes
\begin{equation}
    E_{m}^{-}(f) \sim \frac{\cos\left(\frac{\pi}{4} - \frac{2}{3} |z|^{3/2}\right)}{|z|^{1/4}}
\end{equation}
where $z<0$. Comparing this to the SPA in  Eq.~\eqref{eq:Em-spa}, we find remarkable similarity in the functional form of these approximations. This implies that the SPA and Airy approximations are actually asymptotic expansions with an overlapping region of validity. For approximations that have this type of behavior, the creation of a uniform (or matched) asymptotic expansion is possible using asymptotic matching~\cite{Bender}.

\subsection{Matched Asymptotic Expansion}
\label{sec:matched}

We now consider the creation of a matched asymptotic expansion to approximate Eq.~\eqref{eq:Em-int}. The formal details of the method can be found, for example, in~\cite{Bender,Verhulst,Dingle}, while some applications within the field of gravitational physics can be found in~\cite{Compere:2021zfj,Nakano:2016klh,Ireland:2015cjj,Dixon:2013lva,Mundim:2013vca,Gallouin:2012kb,Yunes:2005nn,Yunes:2006iw,Poujade:2001ie}. In our case, the two regimes of validity have been detailed in Sec.~\ref{sec:spa} \&~\ref{sec:fold}, and the matching region is the region around $f_{\rm max}$. Below, we provide the details of the matching procedure, and include subdominant effects due to the finite limits of integration in Eq.~\eqref{eq:Em-int}.

\subsubsection{A Leading Order Approximation}

Based on the asymptotic behavior of $E_{m}^{-}(f)$ given in Eqs.~\eqref{eq:Em-spa} \&~\eqref{eq:Em-airy}, we propose that a suitable matched asymptotic expansion (MAE) across the critical point at $f=f_{\rm max}$ is
\begin{equation}
    \label{eq:Em-uae}
    \left[E_{m}^{-}(f)\right]_{\rm MAE} = \frac{2\pi}{\sigma^{1/3}\Omega}  e^{\alpha(f)} {\rm Ai}\left[-\beta_{m}(f)\right] e^{2\pi i f t_{p}}\,,
\end{equation}
where $[\alpha(f), \beta_{m}(f)]$ are unknown functions that will be fixed via matching. To do the matching, we define a new variable $\zeta$ such that,
\begin{align}
    f = \frac{m \Omega}{2\pi} \left(1 + e - 2 e \zeta\right)^{2}\,,
\end{align}
which maps the domain $f\in [f_{\rm min}, f_{\rm max}]$ to $\zeta \in [1,0]$. Near the fold catastrophe, $\zeta \ll 1$ and it suffices to consider Taylor expansions of all relevant quantities about $\zeta=0$. Thus, we posit
\begin{align}
    \label{eq:alpha-beta}
    \alpha(f) = \sum_{k=1}^{\infty} \alpha_{k} \zeta^{k}\,, \qquad \beta_{m}(f) = \sum_{k=1}^{\infty} \beta_{k} \zeta^{k}\,.
\end{align}
with unknown coefficients $(\alpha_{k}, \beta_{k})$. The goal of the computation is to determine the coefficients $(\alpha_{k}, \beta_{k})$. 

Typically, when constructing matched asymptotic expansions, one has to match the master function in both regimes of validity. However, by virture of our choice in Eq.~\eqref{eq:Em-uae}, we have already used knowledge of the high frequency expansion in Eq.~\eqref{eq:Em-fold}. As a result, the unknown constants $(\alpha_{k},\beta_{k})$ can all be fixed by performing the matching in the region $f<f_{\rm max}$, where the SPA is the leading order approximation. Performing the asymptotic expansion of Eq.~\eqref{eq:Em-uae} about $\beta \rightarrow \infty$, we have to leading order
\begin{align}
    \label{eq:Em-left}
    \left[E_{m}^{-}(f)\right]_{\rm MAE} &\sim \frac{2\sqrt{\pi}}{\sigma^{1/3} \Omega} \frac{e^{\alpha(f)}}{\left[\beta_{m}(f)\right]^{1/4}} 
    \nn \\
    &\times \cos\left\{\frac{\pi}{4} - \frac{2}{3} \left[\beta_{m}(f)\right]^{3/2}\right\} e^{2\pi i f t_{p}}\,.
\end{align}
We only carry out the expansion here to leading order, since the SPA only constitutes the leading order asymptotic expansion of $E_{m}^{-}(f)$ in the region $f_{\rm min} < f < f_{\rm max}$. Comparing Eq.~\eqref{eq:Em-left} to the SPA in Eq.~\eqref{eq:Em-spa}, one can see that the functions $[\alpha(f),\beta_{m}(f)]$ map directly to the amplitude and phase of the SPA, specifically
\begin{align}
    \label{eq:alpha-matched}
    \alpha(f) &= \ln\left[\sigma^{1/3} \Omega \sqrt{\frac{2}{\ddot{\Psi}_{-,\star}}} \left(\frac{3}{2} \Psi_{-,\star}\right)^{1/6}\right]\,,
    \\
    \label{eq:beta-matched}
    \beta_{m}(f) &= \left(\frac{3}{2} \Psi_{-,\star}\right)^{2/3}\,.
\end{align}
Thus, once one knows the Taylor expansions of $[\Psi_{-,\star}, \ddot{\Psi}_{-,\star}]$ in Eqs.~\eqref{eq:spa-phase} \&~\eqref{eq:spa-amp} about $\zeta = 0$, one can easily map these to Eq.~\eqref{eq:alpha-beta} to obtain $[\alpha_{k}, \beta_{k}]$. We provide these mappings explicitly in Appendix~\ref{app:expand}. At this stage, all of the unknown quantitites are fixed, and the development of the MAE in Eq.~\eqref{eq:Em-uae} is complete.

Before continuing, it is worth noting a few things about these results. First, up to the overall factor of $e^{2\pi i f t_{p}}$, the SPA in Eq.~\eqref{eq:Em-spa}, the high frequency approximation in Eq.~\eqref{eq:Em-airy}, and the MAE in Eq.~\eqref{eq:Em-uae} are all real-valued, while the original integral in Eq.~\eqref{eq:Em-int} appears complex. However, the real part of Eq.~\eqref{eq:Em-int} is actually even on the domain $\ell \in [-\pi,\pi]$, while the imaginary part is odd. As a result, the imaginary part vanishes upon evaluating the integral, and $E_{m}^{-}(f)$ becomes real-valued.

Second, we have not repeated these computations for $E_{m}^{+}$ at this point. As stated previously, the reason for this is that there are no stationary points for $E_{m}^{+}$ for $f > 0$, nor does the saddle point at $V=0$ dominate the integral for $f>f_{\rm max}$. In fact, rather than having the behavior shown in the bottom panels of Fig.~\ref{fig:Em-plot}, the integral for $E_{m}^{+}$ has the opposite behavior, i.e. it becomes highly oscillatory near $V=0$ while oscillating less rapidly near $V=\pm \pi$. As a result, the analysis of $E_{m}^{+}(f)$ becomes simpler when $f>0$, and we provide a suitable analytic approximation in Appendix~\ref{app:Em_plus}. However, for negative frequencies, $E_{m}^{+}(f)$ and $E_{m}^{-}(f)$ switch roles. In fact, it is straightforward to show from Eq.~\eqref{eq:Em-int} that $E_{m}^{+}(-f) = \left[E_{m}^{-}(f)\right]^{\dagger}$, where $\dagger$ corresponds to complex conjugation. 

\subsubsection{Beyond Leading Order}
\label{sec:blo}

The response of the integral in Eq.~\eqref{eq:Em-int} is oscillatory due to the finite limits of integration, while the MAE in Eq.~\eqref{eq:Em-uae} is not since we have performed the asymptotic approximation of taking the limits to infinity. The MAE thus constitutes the leading order terms in an asymptotic expansion of Eq.~\eqref{eq:Em-int}, with the oscillations arising from sub-dominant effects. These oscillatory effects are important to some of the applications in Sec.~\ref{sec:apps}, so we will here provide analytic expressions for these corrections.

To begin, we re-write Eq.~\eqref{eq:Em-int} as
\begin{align}
    \label{eq:int-new}
    E_{m}^{-}(f) = {\cal{F}}\left[e^{-imV}\right] - e^{2\pi i f t_{p}} \left[ {\cal{R}}(f) + {\cal{R}}^{\dagger}(f) \right]
\end{align}
where ${\cal{F}}$ is the (all-time) Fourier transform
\begin{equation}
    \label{eq:fourier}
    {\cal{F}}[g] = \int_{-\infty}^{\infty} dt \; g(t) e^{2\pi i f t}\,,
\end{equation}
with $g(t)$ an arbitrary function, and ${\cal{R}}(f)$ is the remainder integral, defined as
\begin{equation}
    \label{eq:remainder-def}
    {\cal{R}}(f) = \int_{\pi}^{\infty} \frac{d\ell}{n} e^{-imV} e^{2\pi i f \ell/n}\,.
\end{equation}
The Fourier transform in Eq.~\eqref{eq:fourier} is approximated by the MAE in Eq.~\eqref{eq:Em-uae}. Much like the development of the MAE, we must understand the critical points of the integrand of ${\cal{R}}(f)$. Since we are neglecting radiation reaction, the phase is oscillatory in the range $\ell \in [-\pi,\pi]$, and repeats for values outside of this range. However, because we are interested in the response over a single orbit, we only consider the critical points in this range. Thus, the only relevant critical points are the stationary points defined in Eq.~\eqref{eq:Vsp}, and the saddle points at $V=0$ and $V=\pm \pi$. When $f > f_{\rm min}$, the remainder integral is dominated by the saddle points at $V=\pm \pi$. Note that the contribution from $V=-\pi$ is already handled by ${\cal{R}}^{\dagger}(f)$, so it suffices to only consider one of these points.

Expanding the phase of Eq.~\eqref{eq:remainder-def} about $V = \ell = \pi$, we obtain
\begin{equation}
    \label{eq:remainder-approx}
    {\cal{R}}(f) \sim e^{i\Psi_{-,\pi}} \int_{\pi}^{\infty} \frac{d\ell}{n} e^{i \left[\dot{\Psi}_{-,\pi} (\ell - \pi) + \dddot{\Psi}_{-,\pi} (\ell - \pi)^{3}/3!\right]}
\end{equation}
where 
\begin{align}
    \Psi_{-,\pi} &= \frac{2\pi^{2}f}{(1-e^{2})^{3/2}\Omega} - \pi m\,,
    \\
    \dot{\Psi}_{-,\pi} &= \frac{2\pi(f-f_{\rm min})}{n}\,,
    \\
    \dddot{\Psi}_{-,\pi} &= -\frac{2me(1-e)^{5}}{(1-e^{2})^{3/2}}\,.
\end{align}
The integral in Eq.~\eqref{eq:remainder-approx} is a special case of the incomplete Airy function~\cite{IncompleteAiry}, and can be evaluated either by repeated integration by parts, or by Watson's lemma~\cite{Bender} upon suitable deformation of the integration contour. However, it is actually possible to obtain an exact answer to the integral in Eq.~\eqref{eq:remainder-approx} by combining it with ${\cal{R}}^{\dagger}(f)$. The sum ${\cal{R}}(f) + {\cal{R}}^{\dagger}(f)$ can be re-arranged to obtain
\begin{align}
    \label{eq:rem-sum}
    {\cal{R}}(f) + {\cal{R}}^{\dagger}(f) &= \frac{2}{n} \cos\Psi_{-,\pi} \int_{\pi}^{\infty} d\ell \cos\left[\delta \Psi_{\pi}(\ell)\right]
    \nn \\
    &+ \frac{2}{n} \sin\Psi_{-,\pi} \int_{\pi}^{\infty} d\ell \sin\left[\delta \Psi_{\pi}(\ell)\right]
\end{align}
where
\begin{equation}
    \delta \Psi_{\pi}(\ell) = \dot{\Psi}_{-,\pi} \left(\ell - \pi\right) + \frac{\dddot{\Psi}_{-,\pi}}{3!} \left(\ell - \pi\right)^{3}
\end{equation}
By suitable change of variables, the first integral above becomes the integral definition of the Airy function ${\rm Ai}(x)$, while the second becomes the integral definition of Scorer's function ${\rm Gi}(x)$~\cite{Olver:1997:ASF,Abromowitz,NIST:DLMF}, specifically
\begin{equation}
    {\rm Gi}(x) = \frac{1}{\pi} \int_{0}^{\infty} dt \; \sin(x t + t^{3}/3)
\end{equation}
Thus,
\begin{align}
    \label{eq:rem-exact}
    {\cal{R}}(f) + {\cal{R}}^{\dagger}(f) &= \frac{2\pi}{n} \rho(e,m) \bigg\{\cos\Psi_{-,\pi} {\rm Ai}[\gamma(f)] 
    \nn \\
    &+ \sin\Psi_{-,\pi} {\rm Gi}[\gamma(f)]\bigg\}
\end{align}
where
\begin{align}
    \rho(e,m) = \left(\frac{2}{|\dddot{\Psi}_{-,\pi}|}\right)^{1/3}\,, \qquad \gamma(f) = |\dot{\Psi}_{-,\pi}| A(e,m)\,.
\end{align}

The approximation in Eq.~\eqref{eq:rem-exact} holds for $f\ge f_{\rm max}$, but what are the proper approximations for the low frequency regime $f \le f_{\rm min}$ and for the SPA interval $f \in [f_{\rm min}, f_{\rm max}]$? When $f \le f_{\rm min}$, there are no critical points and the inflection points at $V=\pm \pi$ dominate the response. As a result, the integral in Eq.~\eqref{eq:Em-int} for $f \le f_{\rm min}$ can be mapped into the form of Eq.~\eqref{eq:rem-sum}, but with $\Psi_{-,\pi} \rightarrow - \Psi_{-,\pi}$. Meanwhile, in the critical region, the response is dominated by the SPA of Eq.~\eqref{eq:fourier} which results in the MAE. This does not hold for $f < f_{\rm min}$, and thus, must be suitable windowed. The form of the SPA in Eq.~\eqref{eq:Em-spa} results from taking the limits of integration to infinity, but only holds up to a remainder of ${\cal{O}}(|\ddot{\Psi}_{-,\star}|^{-1})$. These corrections actually result from the fact that the original integral is only over a finite time window. Further, these effects are actually subdominant compared to those of Eq.~\eqref{eq:rem-exact}, with the exception of a small region near $f=f_{\rm min}$. Thus, a suitable approximant to Eq.~\eqref{eq:Em-int} is
\begin{align}
    \label{eq:Em-final}
    E_{m}^{-}(f) &\sim e^{2\pi i f t_{p}}  \Bigg(\left[E_{m}^{-}(f)\right]_{\rm UAE} \Theta\left(f-f_{\rm min}\right) 
    \nn \\
    &+ \frac{2\pi}{n} A(e,m) \bigg\{\cos\Psi_{-,\pi} {\rm Ai}\left[B(f)\right] 
    \nn \\
    &+ {\rm sign}(f-f_{\rm min}) \sin\Psi_{-,\pi} {\rm Gi}\left[B(f)\right]\bigg\}\Bigg)\,,
\end{align}
where $\Theta(x)$ is the Heaviside step function. This will constitute our final asymptotic expansion of Eq.~\eqref{eq:Em-int}.

\begin{figure*}[hbt!]
    \centering
    \includegraphics[width=\textwidth, trim={4cm 1cm 4cm 0cm}, clip]{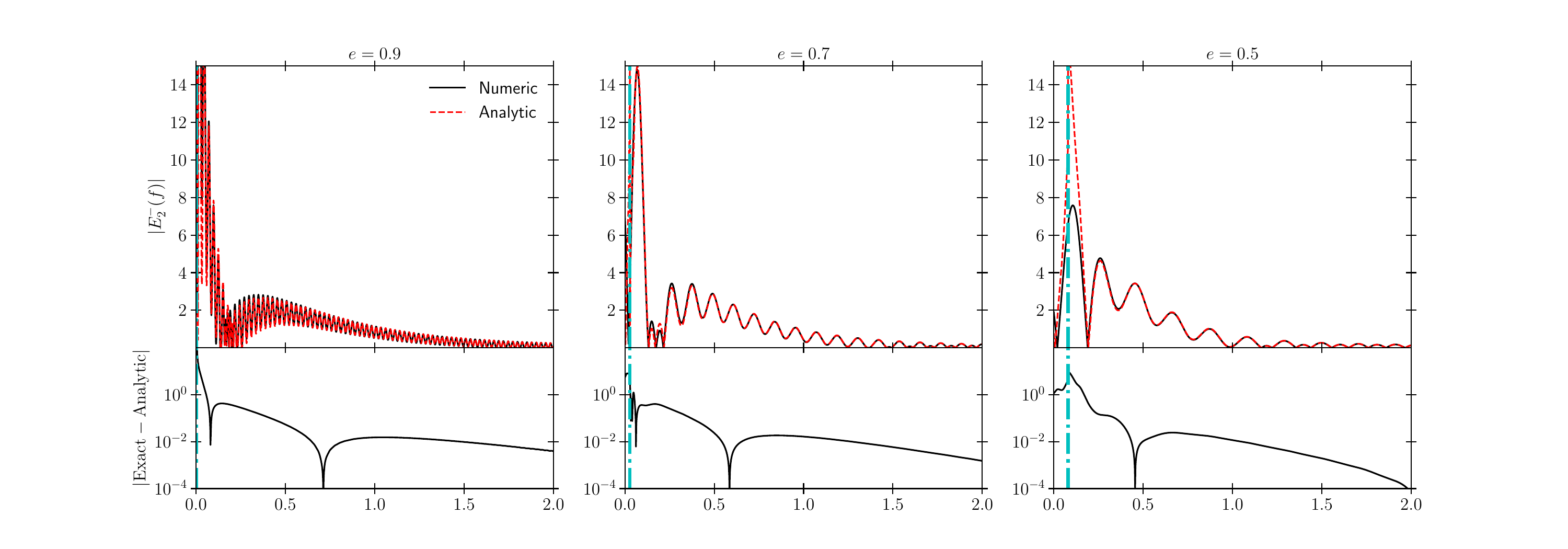}
    \includegraphics[width=\textwidth, trim={4cm 1cm 4cm 2cm}, clip]{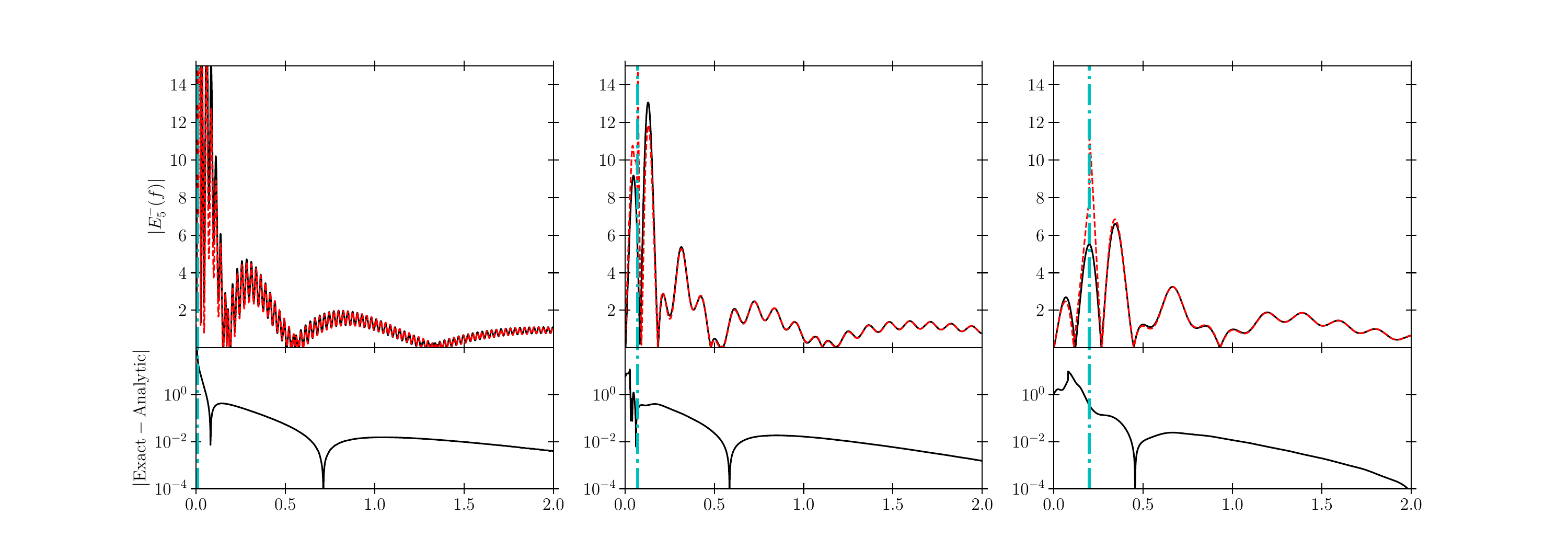}
    \includegraphics[width=\textwidth, trim={4cm 0cm 4cm 2cm}, clip]{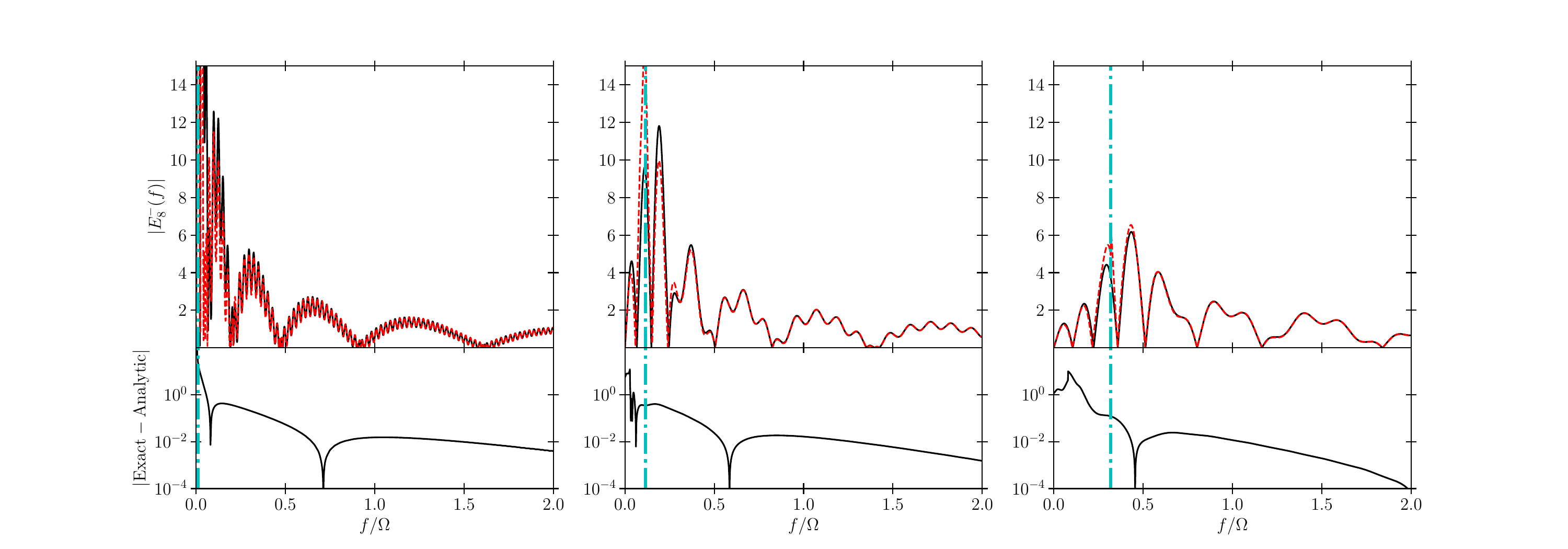}
    \caption{\label{fig:Em-comp} Comparison of the closed form analytic approximation for $E_{m}^{-}(f)$ in Eq.~\eqref{eq:Em-final} (dashed red line) to a numerical computation of Eq.~\eqref{eq:Em-int} (solid black line) for $e = [0.9, 0.7, 0.5]$ (left, middle, right columns) and $m = [2, 5, 8]$ (top, middle, and bottom rows). The bottom panel of each plot provides the difference between the exact (numerical) result and the analytic approximation. The vertical dot-dashed lines provide the value of $f_{\rm min}$ given in Eq.~\eqref{eq:fmin-max} for each case.}
\end{figure*}

Fig.~\ref{fig:Em-comp} provides a comparison of the analytic approximation in Eq.~\eqref{eq:Em-final} to the numerical computation of Eq.~\eqref{eq:Em-int} as a function of the frequency $f$. The numerical calculation is achieved by performing a change of variables from time $t$ to the true anomaly $V$ by using Eq.~\eqref{eq:Vdot} in Eq.~\eqref{eq:Em-int}. The benefit of this step is that it does not require us to perform a numerical integration of the orbit to obtain $V(t)$ numerically. After this change of variable, we sample the integrand in Eq.~\eqref{eq:Em-int} with $2^{24}$ points, and approximate the integral by the summation over these samples. The total number of points is chosen to obtain sufficient accuracy to properly determine the difference between the numerical and analytic results. One could choose a finer sampling than what was chosen here, but this requires increased computation time.

Each column in Fig.~\ref{fig:Em-comp} represents different values of the eccentricity, specifically $[0.9, 0.7, 0.5]$ (left, middle, right), while each row corresponds to different $m$ values, specifically $[2,5,8]$ (top, middle, bottom). The bottom panel of each plot provides the difference between the numeric and analytic result, providing an estimate of error in the analytic approximations for a given frequency. The vertical dot-dashed line provides the value of $f_{\rm min}$ for each case. The approximant of Eq.~\eqref{eq:Em-final} generally models Eq.~\eqref{eq:Em-int} well, but the errors typically become large in two cases: when $f \sim f_{\rm min}$ and when $e \rightarrow 1$. The first of these is a result of the asymptotic matching being performed at $f=f_{\rm max}$, and thus, the approximations will become less accurate as the frequency approaches $f_{\rm min}$. The latter results from the fact that both the frequency of oscillations and amplitude of Eq.~\eqref{eq:Em-int} increase as $e\rightarrow 1$ and $f \rightarrow f_{\rm min}$. Hence, small errors resulting from the approximations used to obtain Eq.~\eqref{eq:Em-final} will generally diverge as one approaches the parabolic limit. We will discuss practical issues resulting from this in Sec.~\ref{sec:apps}. 

\section{Applications}
\label{sec:apps}

Having completing our methodology for approximating Eq.~\eqref{eq:Em-int}, we now turn our attention to a few applications of the approximations developed in the previous section. The applications presented here are only a small subset which are relevant to the study of GWs from eccentric binaries, particularly those formed through dynamical capture interactions.

\subsection{Effective Fly-By Waveforms}
\label{sec:efb}

One of the present challenges of GW modeling is the creation of waveforms that accurately model the high eccentricity regime, where the GWs are characterized by bursts emitted during periastron passage. Some recent work toward this are the EOB waveforms of~\cite{RamosBuades:2020:EccentricSearchEfficiency,Nagar:2021:EccentricWaveform}, and the EFB waveforms of~\cite{Loutrel:2019kky,Loutrel:2020jfx}. We here show that the methods of Sec.~\ref{sec:cat} allow for a simplified development of EFB waveforms, compared to the re-summation procedures presented in~\cite{Loutrel:2019kky}. For simplicity, we neglect the effect of radiation reaction, but discuss how to properly implement it, as well as other PN corrections, later in this section.

The analysis of Sec.~\ref{sec:cat} was carried out to Newtonian order. At this PN order, the GW polarizations are described by the quadrupole formula~\cite{PoissonWill}, and in the time domain are~\cite{Moreno-Garrido,Moore:2018kvz}
\begin{align}
    \label{eq:h-pc}
    h_{+,\times}(t) &= -\frac{2\eta M^{2}}{p D_{L}} \sum_{m=-3}^{3} A_{+,\times}^{(m)}(e,\iota,\beta) e^{imV(t)}
\end{align}
where $\eta$ is the binary's symmetric mass ratio, $\iota$ is the binary's inclination angle relative to the line of sight, and $\beta$ is an arbitrary polarization angle. The EFB approach provides an approximation for the waveforms by treating each orbit that creates a GW burst as a fly-by rather than a repeating elliptical orbit. For the application of the methods in Sec.~\ref{sec:cat}, this amounts to performing a Fourier transform of Eq.~\eqref{eq:h-pc}, suitably windowed over a single orbit. For a numerical computation of the Fourier domain waveform, this simply becomes the computation of the FFT for Eq.~\eqref{eq:h-pc} over a single orbit. To analytically approximate the Fourier transform, one simply has to apply Eq.~\eqref{eq:int-new}. As a result, this simply amounts to taking $e^{imV} \rightarrow E_{m}^{{\rm sign}(m)}(f)$ in Eq.~\eqref{eq:h-pc}, i.e. 
\begin{equation}
    \label{eq:h-efb}
    \tilde{h}_{+,\times}^{\rm EFB}(f) = -\frac{2\eta M^{2}}{p D_{L}} \sum_{m=-3}^{3} A_{+,\times}^{(m)}(e,\iota,\beta) E_{m}^{{\rm sign}(m)}(f)\,,
\end{equation}
with each $E_{m}^{{\rm sign}(m)}(f)$ given analytically by Eq.~\eqref{eq:Em-final}.

To test the accuracy of the approximations used to obtain Eq.~\eqref{eq:Em-final} for $E_{m}^{-}(f)$ and Eq.~\eqref{eq:Ep-approx} for $E_{m}^{+}(f)$, we compute the match
\begin{equation}
    \label{eq:match}
    {\cal{M}} = \underset{t_{p}}{\max} \langle \hat{h}^{\rm FFT} | \hat{h}^{\rm EFB} \rangle
\end{equation}
where $\langle \; | \; \rangle$ is the inner product between waveforms, defined as
\begin{equation}
    \label{eq:ip}
    \langle A | B\rangle = \frac{4}{5} {\rm Re} \int_{f_{\rm low}}^{f_{\rm high}} df \left[ A_{+}(f) B_{+}^{\dagger}(f) + A_{\times}(f) B^{\dagger}_{\times}(f)\right]\,,
\end{equation}
where ${\rm Re}$ corresponds to the real part of the expression, $\dagger$ corresponds to complex conjugation, and $\hat{A} = A/\sqrt{\langle A | A\rangle}$. The inner product in Eq.~\eqref{eq:ip} is analogous to the sky-averaged, noise-weighted inner product often used is mock analysis studies of waveforms, albeit for a detector with white noise~\footnote{The detector response to a GW with polarizations $h_{+/\times}(t)$ is $h(t)=F_{+}(\alpha,\delta,\psi) h_{+}(t) + F_{\times}(\alpha, \delta, \psi) h_{\times}(t)$, where $(\alpha,\delta,\psi)$ are the right ascension and declination of the source in the detector frame, and $\psi$ is the polarization angle. To obtain the connection between the standard noise-weighted inner product (see, for example, Eq. (5.1) in~\cite{Buonanno:2009zt}) and Eq.~\eqref{eq:ip}, one simply has to compute the sky- and polarization-averaged beam pattern functions, specifically $\langle F_{+}^{2} \rangle = 1/5 = \langle F_{\times}^{2} \rangle$ and $\langle F_{+} F_{\times} \rangle = 0$.}. We don't consider individual detectors here since the waveforms in Eq.~\eqref{eq:h-efb} are not representative of waveforms we might expect from nature, owing to the fact that they are Newtonian order and neglect radiation reaction. Further, we want the results of the match analysis to be detector agnostic, hence why we choose white noise instead of detector specific noise. The range of values the match can take are $[0,1]$, and in this context, the closer the match is to unity, the more accurate the EFB waveform is to the numerical waveform.

\begin{figure}[hbt]
    \centering
    \includegraphics[width=\columnwidth]{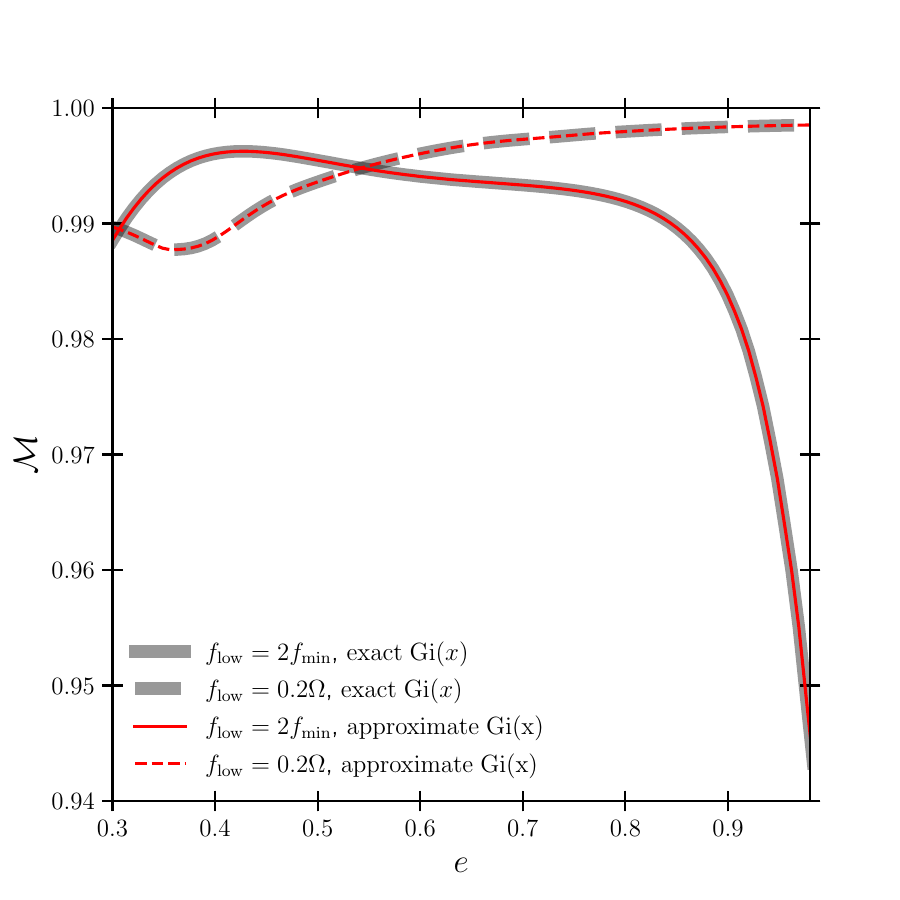}
    \caption{Match ${\cal{M}}$ between numerical FFT waveforms and the analytic EFB waveforms of Eq.~\eqref{eq:h-efb}. Solid lines correspond to cases where the lower limit of integration in Eq.~\eqref{eq:ip} is taken to be $f_{\rm low} = 2f_{\rm min}$, while dashed lines take $f_{\rm low} = 0.2\Omega$. The match is generally greater than 0.98, except in the high eccentricity cases when $f_{\rm low} = 2f_{\rm min}$, due to error in the analytic approximations near $f_{\rm min}$. Gray lines correspond to cases when the Scorer's function in Eq.~\eqref{eq:Em-final} is computed using the \texttt{mpmath} module in \texttt{Python}, while red lines use the approximation detailed in Appendix~\ref{app:gi}. The approximation does not result in a significant loss of the match.}
    \label{fig:match}
\end{figure}

From a practical standpoint, computing Eq.~\eqref{eq:Em-final} can be difficult due to its dependence on special functions. While Airy functions are well documented, Scorer's functions are not, and in Python, are only implemented numerically in the \texttt{mpmath} package~\cite{mpmath}. While this isn't a problem in terms of evaluation, it does slow down the evaluation of Eq.~\eqref{eq:Em-final} due to the arbitrary precision nature of \texttt{mpmath}. To speed this up, we approximate Scorer's function via the method in Appendix~\ref{app:gi}, which is roughly one hundred times faster to evaluate than the implementation in \texttt{mpmath}, but has the drawback of only being approximate.

Due to neglecting radiation, the maximization over $t_{p}$ in Eq.~\eqref{eq:match} is trivially given by $t_{p} = -T_{\rm orb}/2 = \pi/\Omega(1-e^{2})^{3/2}$. When performing the integral in Eq.~\eqref{eq:ip}, we set $f_{\rm high} = 3\Omega$ and allow $f_{\rm low}$ to vary. The former is due to the fact that the integrand decays exponentially for $f\gg \Omega$, and the match is insensitive to values above this choice of the upper limit of integration. For the latter, we seek to test the accuracy of the approximant in Eq.~\eqref{eq:Em-final} as a function of an artificial cutoff at low frequency. As a result, we compute the match for two cases: when $f_{\rm low} = 2 f_{\rm min}$ and $f_{\rm low} = 0.2\Omega$. 

The results of the match computation are plotted in Fig.~\ref{fig:match} as a function of $e$ and for $\eta=1/4$. The solid lines correspond to the case when $f_{\rm low} = 2 f_{\rm min}$, while the dashed lines correspond to $f_{\rm low} = 0.2\Omega$. In the former case, the match drops off rapidly as $e\rightarrow 1$, for the same reasons explained below Eq.~\eqref{eq:Em-final}, specifically the limited accuracy of the approximation near $f=f_{\rm min}$. The errors in the approximations used to obtain Eq.~\eqref{eq:Em-final} coupled to the rapidly oscillating response of Eq.~\eqref{eq:Em-int} in the high eccentricity limit results in significant dephasing between the analytic and numerical waveforms at low frequencies. Thus, from the practical standpoint of using Eq.~\eqref{eq:Em-final} for eccentric burst waveforms, one must be cautious about choosing a suitable low frequency cutoff, or alternatively, append Eq.~\eqref{eq:Em-final} with a suitable approximation in the region around $f = f_{\rm min}$. It is worth noting however, that we are using white noise for this computation, whereas realistic ground-based detectors are less sensitive at low frequencies due to seismic noise~\cite{Aasi_2015}, which may aleviate some of the problems with the low frequency cutoffs of the waveforms. A more thorough analysis of this low frequency cutoff with realistic, higher PN order waveforms and detector noise will be carried out in future work.

In addition, we also compute the match between two different sets of EFB waveforms, specifically those which use the $\texttt{mpmath}$ package to evaluate ${\rm Gi}(x)$ (gray lines), and those that use the approximation in Appendix~\ref{app:gi} (red lines). The match doesn't change significantly between these two cases, and thus, the fast approximation of Appendix~\ref{app:gi} should be favored over the more accurate, but slower, computation.

Before moving onto the next application, we will briefly discuss the inclusion of higher PN order effects in the waveforms. Generally, the equations of motion of the PN two body problem take the form~\cite{blanchet-lrr}
\begin{equation}
    \label{eq:eom}
    \vec{a} = - \frac{M}{r^{2}}\vec{n} + \delta \vec{f}_{\rm cons} + \delta \vec{f}_{\rm diss}
\end{equation}
where $\vec{a}$ is the relative acceleration, $\vec{n}$ is the relative unit normal vector, $\delta f_{\rm cons}$ are the conservative PN corrections, and $\delta \vec{f}_{\rm diss}$  are the dissipative PN corrections due to radiation reaction. To lowest PN order,
\begin{align}
    \delta \vec{f}_{\rm cons} &= \vec{f}_{\rm 1PN} + {\cal{O}}(c^{-6})\,,
    \\
    \delta \vec{f}_{\rm diss} &= \vec{f}_{\rm 2.5PN} + {\cal{O}}(c^{-9})\,,
\end{align}
where $\vec{f}_{n{\rm PN}}$ are the n-th PN order corrections to the relative force, and $c$ is the speed of light. Arguably, the most powerful method for solving these equations is the method of osculating orbits~\cite{PoissonWill,LincolnWill,Mora:2003wt}. The Newtonian two-body problem admits the solution $V = V(t,\mu^{a})$ and $r = r[V(t),\mu^{a}]$, with constants of motion $\mu^{a}$. The method of osculating orbits promotes $\mu^{a}$ to functions of time, which satisfy the osculating equations
\begin{equation}
    \frac{d\mu^{a}}{dt} = {\cal{F}}\left[V(t), \mu^{b}(t)\right]\,,
\end{equation}
where the ${\cal{F}}^{a}$ depend on the components of $\delta \vec{f}_{\rm cons/diss}$. 

Generally, the osculating equations are non-linear and do not admit an exact, closed-form solution. The most common method of solving them is to employ multiple scale analysis~\cite{Bender}, but in the context of approximating Eq.~\eqref{eq:Em-int}, a simpler method would be to perturb about the values of $\mu^{a}$ at pericenter passage, and truncate the expansion at highest PN order taken in Eq.~\eqref{eq:eom}~\cite{Arredondo:2021rdt}. The analysis carried out in Sec.~\ref{sec:cat} still holds, but now the stationary points will drift due to PN corrections. We do not perform that analysis here for two reasons. First, to obtain the most accurate models of GW bursts from eccentric binaries, one will need the higher PN order amplitude corrections to Eq.~\eqref{eq:h-pc}, which at sufficiently high PN order will contain the hereditary tail and memory contributions~\cite{Mishra:2015bqa,Boetzel:2019nfw,Ebersold:2019kdc}. Computing these, especially the latter, for highly eccentric binaries goes outside of the scope of this work. Second, the EFB waveforms require a timing model that accurately tracks the time of pericenter passages~\cite{Loutrel:2019kky,Arredondo:2021rdt}, in a similar way to pulsar timing models, and also goes outside of the scope of this work. We plan to address both of these points in future work.

\subsection{From Waves to Bursts}
\label{eq:trans}

One of the earliest predictions from the study of GWs within PN theory was the so-called circularization of an inspiraling binary system, i.e. the GWs cause the orbital eccentricity to decay until it becomes negligibly small and the binary enters into a plunge state in the final few orbits~\cite{Peters:1963ux, Peters:1964zz}. Due to circularization, the GWs must transition from burst-like behavior to wave-like behavior as the eccentricity decays. We here show that there is a means of understanding when this transition occurs using the methods of the previous Sec.~\ref{sec:cat}.

How does one actually quantify when this transition occurs? One may be tempted to use a notion of the pericenter passage timescale, such as Eq.~(12) in~\cite{Turner}, and it's relation to the orbital period to answer this. However, for any given eccentricity $e$, the pericenter passage timescale is always shorter than the orbital timescale, so this does not provide any useful information. Another temptation that may arise is to move completely into the frequency domain via Fourier transforms defined in Eq.~\eqref{eq:fourier}. Much of GW data analysis is done in Fourier space, and GW bursts from highly eccentric binaries are known to have a characteristic high frequency tail, which is approximated by the high frequency response of $E_{m}^{-}(f)$ in Eq.~\eqref{eq:Em-airy}. However, by virtue of the limits of integration in Eq.~\eqref{eq:fourier}, all time information is integrated out of the function under consideration (the GWs in this case). This is detrimental when trying to consider the problem at hand, since the secular behavior of the eccentricity directly maps to time. For example, within the quadrupole approximation and assuming adiabaticity, the eccentricity evolves according to~\cite{Peters:1964zz}
\begin{equation}
    \frac{de}{dt} = -\frac{304}{15} e \frac{\eta}{M} \left(\frac{M}{p}\right)^{4} \left(1-e^{2}\right)^{3/2} \left(1 + \frac{121}{304} e^{2}\right)\,.
\end{equation}
The solution is to use short-time Fourier transforms like the one defined in Eq.~\eqref{eq:Em-int}, or more general wavelet transforms~\cite{TorrenceCompo}, since these contain both time and frequency information. 

At leading PN order, the GWs polarizations are given in Eq.~\eqref{eq:h-pc} in terms of harmonics of the true anomaly $V$, with the $m=2$ harmonic being dominant since it's amplitude is not coupled to the eccentricity. The calculation then reduces down to computing Eq.~\eqref{eq:Em-int}. It is useful at this stage to understand the low eccentricity behavior of Eq.~\eqref{eq:Em-int}, and as a result, the waveforms. For $e \ll 1$, Eq.~\eqref{eq:Vdot} can be solve perturbatively in $e$ to obtain,
\begin{equation}
    V = \ell + 2 e \sin \ell + {\cal{O}}(e^{2})\,.
\end{equation}
Then, up to linear order in $e$,
\begin{equation}
    E_{m}^{-}(f) = \int_{-\pi}^{\pi} \frac{d\ell}{n} e^{2\pi i f t - i m \ell} \left\{1 + 2ie m \sin\ell + {\cal{O}}(e^{2})\right\}\,.
\end{equation}
The response of $E_{m}^{-}(f)$ then depends on the value of $f$. If the frequency is an integer multiple of the orbital frequency, i.e. $f = k n/2\pi$ for any integer $k$, then
\begin{equation}
    \label{eq:Em-smalle}
    E_{m}^{-}(kn/2\pi) = \frac{2\pi}{n} \left[\delta_{k,m} + e m \left(\delta_{k,m+1} - \delta_{k,m-1}\right) + {\cal{O}}(e^{2})\right]\,.
\end{equation}
The response is then dominated by spectral lines at harmonic numbers $m, m+1$, and $m-1$. The subdominant lines are separated from the dominant harmonic by the orbital frequency $F_{\rm orb} = n/2\pi$. This is the equivalent response one expects from GWs in Fourier space for low eccentricities. When $f$ takes any other value, the Kronecker delta's in Eq.~\eqref{eq:Em-smalle} are replaced with sinc functions. 

Now, consider the alternative computation of using the results of Sec.~\ref{sec:cat} to obtain $E_{m}^{-}(f)$ in the small eccentricity limit. Even for small eccentricities, the stationary points that allow for application of the SPA still exist and are still defined by Eq.~\eqref{eq:Vsp}. The frequency interval where the SPA is valid is defined by $f \in (f_{\rm min}, f_{\rm max})$, with $f_{\rm min/max}$ given in Eq.~\eqref{eq:fmin-max}. When $e \sim 0$, both $f_{\rm min} \sim f_{\rm max} \sim m n$, and there is only one harmonic contained in the SPA window. Further $f_{\rm max} - f_{\rm min} \ll n$, and so the region of validity of the SPA is small. However, as the eccentricity increases, $f_{\rm max}$ increases while $f_{\rm min}$ decreases\footnote{It is worth noting that, at leading PN order, these quantities are independent of the mass ratio. At higher PN order, this is no longer true, and will depend weakly on the mass ratio, due to the fact that the corrections are coupled to the orbital velocity, and PN theory assumes $v/c\ll 1$.} . At certain values of the eccentricity, $f_{\rm min/max}$ will becoming larger/smaller than the frequencies of the subdominant harmonics in Eq.~\eqref{eq:Em-smalle}. The first crossing to occur happens when $f_{\rm max} = (m+1)n/2\pi$, which defines the condition for when the response of Eq.~\eqref{eq:Em-int} will no longer be described by the summation of individual orbital harmonics. For GWs, the dominant harmonic of the waveform is $m=2$, and applying this to the condition gives $e_{\rm crit} = 0.191059$.

\begin{figure*}[hbt!]
    \centering
    \includegraphics[width=\textwidth, trim={4cm 1cm 4cm 0cm}, clip]{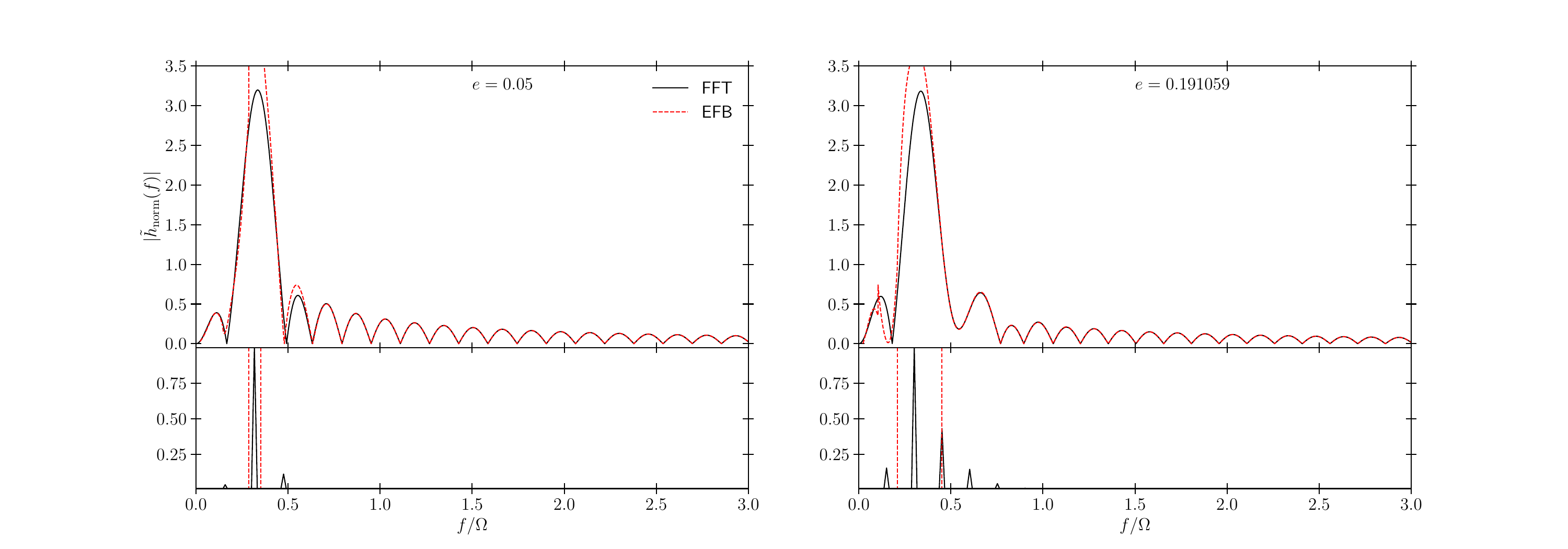}
    \includegraphics[width=\textwidth, trim={4cm 0cm 4cm 0.9cm}, clip]{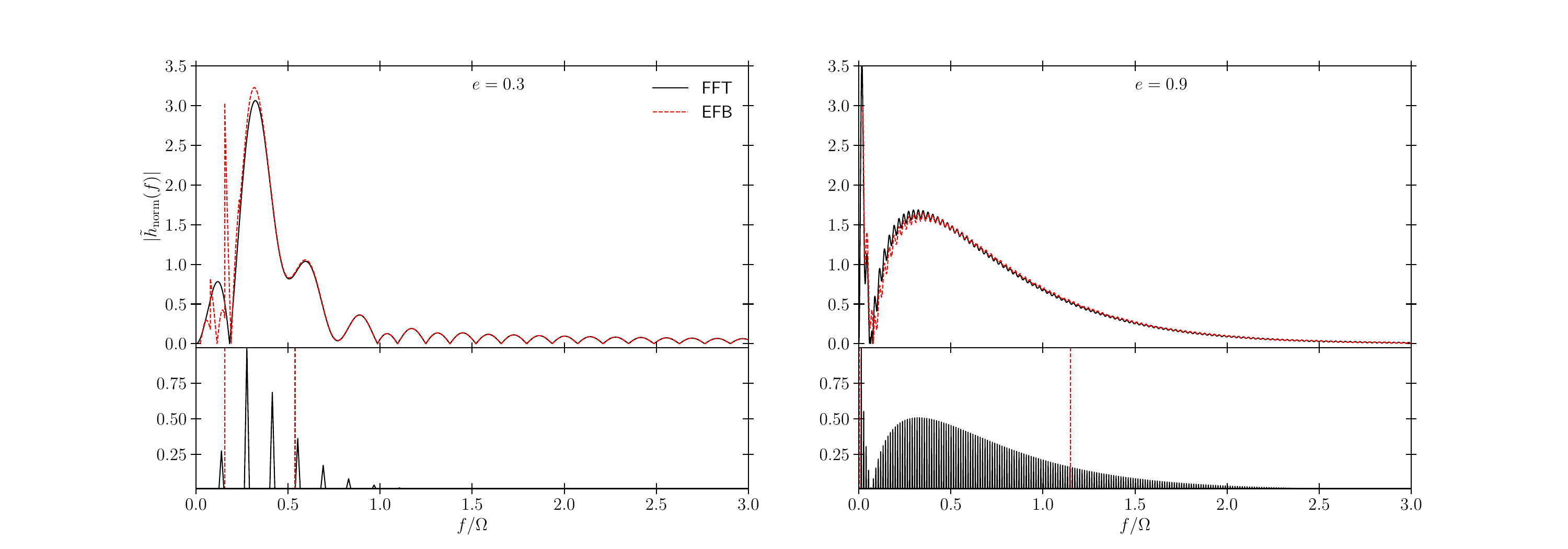}
    \caption{\label{fig:spec} Top panels: Comparisons between FFT waveforms and analytic EFB waveforms for $e=0.05$ (upper left), $e=e_{\rm crit}=0.191059$ (upper right), $e=0.3$ (lower left), and $e=0.9$ (lower right). As the eccentricity increases above $e_{\rm crit}$, the exponential tail that characterizes burst like emission appears. Bottom panels: Plots of the spectral lines corresponding to the orbital harmonics contained in the Newtonian waveforms of Eq.~\eqref{eq:h-pc}. The red dashed lines provide the values of $f_{\rm min/max}$ for each value of $e$, with the region between them defining the SPA window. As the eccentricity increases, the SPA window expands. At $e=e_{\rm crit}$, the third orbital harmonic enters the window, and the tail of the waveform begins to develop. At higher eccentricities, more harmonics enter the window, and the tail extends to higher frequency.}
\end{figure*}

Fig.~\ref{fig:spec} provides an illustrative example of this behavior. The top panel of each plot provides a comparison between the numerical FFT of Eq.~\eqref{eq:h-pc} over one orbit and the EFB approximation in Eq.~\eqref{eq:h-efb}. The values of the waveforms are normalized such that $M/p = 1 = M/D_{L}$. At low eccentricity, for example $e=0.05$ (upper left plot), the waveform is approximately a sinc function due to the finite time window. The bottom panel displays the spectral lines corresponding to the orbital harmonics contained in the waveform, normalized by the value of the maximum orbital harmonic. The dashed lines provide the values of $f_{\rm min}$ and $f_{\rm max}$, which are centered around the second harmonic. As the eccentricity grows to $e=e_{\rm crit}$ (upper right plot), the waveforms begin to develop a high frequency tail, due to the fact that $f_{\rm max} = 3n/2\pi$, and the third orbital harmonic enters the SPA interval. For higher eccentricities, $e=0.3$ (bottom left plot) and $e=0.9$ (bottom right plot), more harmonics are contained in the SPA interval and the tail of the waveform extends to higher frequencies.

The value of the critical eccentricity may seem surprisingly low, since even at $e=e_{\rm crit}$, realistic waveforms do not appear burst-like (see for example~\cite{RamosBuades:2022:EccentricWaveform}). However, it is important to note that the transition from burst-like emission to wave-like emission is not prompt, but continues adiabatically as the eccentricity decays during a binary coalescence. In addition, the analysis carried out here only holds to Newtonian order, and it is likely that PN corrections will modify the value of $e_{\rm crit}$. 

We conclude this section by pointing out an interesting relationship between the properties of bursts that is analogous to those of Bose-Einstein condensates. In the high eccentricity limit, the size of the SPA interval is well approximated by $f_{\rm max}$, since $f_{\rm min}\rightarrow 0$ as $e\rightarrow1$. The number of orbital harmonics contained in the SPA window is then approximated by $N_{\rm harm} = f_{\rm max}/F_{\rm orb}$, where recall $F_{\rm orb}= n/2\pi$. The high frequency tail of the burst is characterized by $\sigma_{f} = \Omega \sigma^{1/3}/2\pi$, where $\sigma$ is given in Eq.~\eqref{eq:sigma-def}. One can then define an effective wavelength $\lambda = 1/\sigma_{f}$, and an effective scale $\lambda_{\rm max} = 1/f_{\rm max}$. Using these definitions, the number of orbital harmonics in the SPA interval is then related to $\lambda$ through
\begin{equation}
    \label{eq:Nharm}
    N_{\rm harm} = 1 - \left(\frac{\lambda_{\rm max}}{\lambda}\right)^{3}\,.
\end{equation}
This expression is equivalent to the average occupation number of the ground state (or condensate) of an ideal Bose gas~\cite{Reif}, with $\lambda$ being recognized as the thermal de Broglie wavelength and $\lambda_{\rm max}$ the critical wavelength. It is worth noting that this connection only applies in the high eccentricity limit, since Eq.~\eqref{eq:Nharm} only applies as $e\rightarrow1$. For lower eccentricities, Eq.~\eqref{eq:Nharm} will be corrected.

\subsection{Hansen Coefficients \& Amplitudes of Dynamical Tides}
\label{sec:hans}

Hansen coefficients provide the Fourier series representation of generic expressions describing Keplerian orbits~\cite{Hansen, HughesHansen}. Specifically, if $r$ is the relative radial separation of the two bodies, and $a = p/(1-e^{2})$ is the semi-major axis of the orbit, the Hansen coefficients are defined as
\begin{align}
    \label{eq:hansen-def}
    X_{k}^{q,m}(e) = \frac{1}{2\pi} \int_{-\pi}^{\pi} d\ell \; \left(\frac{r}{a}\right)^{q} e^{i m V} e^{-ik\ell}\,,
\end{align}
which are the coefficients of the series
\begin{equation}
    \left(\frac{r}{a}\right)^{q} e^{imV} = \sum_{k=-\infty}^{\infty} X_{k}^{q,m}(e) e^{ik\ell}\,.
\end{equation}
In select cases, the $X_{k}^{q,m}(e)$ coefficients are known exactly in terms of hypergeometric functions~\cite{Sadov}, but generically, they are only known in small eccentricity expansions or are computed numerically. The Hansen coefficients are sufficiently generic that they have a wide range of applications, including PN waveforms from eccentric binaries~\cite{Mikoczi:2015ewa}, and the tidal response of stars in eccentric orbits~\cite{Arredondo:2021rdt,Yang:2018bzx,Yang:2019kmf}. To exemplify how the approximations derived herein can be used to approximate Hansen coefficients, we focus on the last of these, specifically on the excitation of f-modes in highly eccentric orbits. 

In~\cite{Arredondo:2021rdt}, it was shown that under a suitable re-summation scheme, the f-mode response takes the form
\begin{align}
    \label{eq:dyn-tide}
    Q_{m}(t) &= \pi \frac{M}{a^{3}} \frac{W_{m} K_{m}}{in\sqrt{\omega^{2} - \gamma^{2}}} e^{-\gamma(t-t_{p})} \left[X^{-3,-m}_{k_{+}} e^{i\sqrt{\omega^{2}-\gamma^{2}}(t-t_{p})}
    \right.
    \nn \\
    &\left.
    - X_{k_{-}}^{-3,-m} e^{-i\sqrt{\omega^{2}-\gamma^{2}}(t-t_{p})} \right]
\end{align}
where $m=0,\pm2$, $K_{m}$ is an equation of state (EOS) dependent parameter defined in Eq.~(12) of~\cite{Arredondo:2021rdt}, $k_{\pm} = (i\gamma \pm \sqrt{\omega^{2} - \gamma^{2}})/n $, $(\omega,\gamma)$ are the frequency and damping coefficient of the mode, and $W_{m}$ is given in Eq.~(24) of~\cite{PressTeukolsky}. Since $k_{\pm}$ are complex and non-integers, the coefficients $X^{-3,-m}_{k_{\pm}}$ are defined by analytic continuation of Eq.~\eqref{eq:hansen-def}. For Keplerian orbits, $r$ is given by Eq.~\eqref{eq:r12}. Expanding out the integrand in Eq.~\eqref{eq:hansen-def}, $X^{-3,-m}_{k}$ for arbitrary $k$ becomes
\begin{align}
    \label{eq:hansen-to-Em}
    X^{-3,-m}_{k} &= \frac{\Omega}{2\pi(1-e^{2})^{3/2}} \sum_{j=-3}^{3} \rho_{j}(e) \left[E_{|j-m|}^{{\rm sign}(j-m)}(\tilde{k}\Omega)\right]^{\dagger}
\end{align}
with $\tilde{k} = k n/2\pi\Omega$,
\begin{align}
    \rho_{0} &= 1 + \frac{3}{2} e^{2}\,, \qquad \rho_{1} = \frac{3}{2} e + \frac{3}{8} e^{3}\,,
    \nn \\
    \rho_{2} &= \frac{3}{4} e^{2}\,, \qquad \rho_{3} = \frac{1}{8} e^{3}\,,
\end{align}
and $\rho_{-j}(e) = \rho_{j}(e)$. 

Can the approximations of Sec.~\ref{sec:matched} be applied to here to approximate the complex Hansen coefficients in Eq.~\eqref{eq:dyn-tide}? We argue that the answer is yes, with a few caveats. From the definition of $E_{m}^{\pm}(f)$ in Eq.~\eqref{eq:Em-int}, take the analytic continuation into the complex plane, i.e. $f \rightarrow \tilde{f} = f_{R} + i f_{I}$, with $f_{R,I}$ the real and imaginary parts. Under this transformation, Eq.~\eqref{eq:Em-int} becomes
\begin{equation}
    \label{eq:Em-mod}
    E_{m}^{\pm}(\tilde{f}) = \int_{-\pi}^{\pi} \frac{d\ell}{n} e^{-2\pi f_{I} t} e^{\pm i m V} e^{2\pi i f_{R} t}\,,
\end{equation}
which is modified by an exponential damping factor. Much of the approximations of Sec.~\ref{sec:cat} rely on the application of Watson's lemma~\cite{Bender}, which requires that the integrand be compactly supported. If this is violated, one cannot take the limits of integration to infinity, as was done in Sec.~\ref{sec:spa} \&~\ref{sec:fold}. The exponential factor in Eq.~\eqref{eq:Em-mod} decays as $t\rightarrow \infty$, but grows when $t\rightarrow -\infty$. It thus seems that we cannot perform this extension of the limits of integration. However, it is not strictly necessary to perform this step when evaluating the integral. It is merely a useful tool for approximating the final result in terms of already known special functions. Further, we point out that the remainder integral from Eq.~\eqref{eq:remainder-def} would necessarily cancel out any divergences that arise from applying this step to Eq.~\eqref{eq:Em-mod}.

A second caveat is associated with the fact that, when applying the approximations of Sec.~\ref{sec:cat}, we did not assume that there was an overall exponential factor (or any amplitude factor) in the integrand of Eq.~\eqref{eq:Em-int}. The response of the integral will change depending on the behavior of any amplitude terms. The stationary points and fold catastrophe of Eq.~\eqref{eq:Em-int} will still exist, however. For the problem at hand, the exponential factor in Eq.~\eqref{eq:Em-mod} will amplify one of the stationary points while suppressing the other. If the amplification/suppression is sufficiently large, than the asymptotic matching of Sec.~\ref{sec:matched} would need to be modified because the contributions to the integral from each stationary point are no longer equal, and the stationary phase approximation cannot be written in the form of Eq.~\eqref{eq:Em-spa}. The caveat is then that the imaginary part of the frequency $f_{I}$ must be sufficiently small as to avoid this issue, or rather ${\rm Im}[\tilde{k}]\ll 1$. For eccentric binaries emitting GWs in the ground-based detection band, this is generally true since the f-mode damping time is typically longer than the orbital period. 

Under the assumption ${\rm Im}[\tilde{k}] \ll 1$, we can approximate the behavior of $E_{m}^{\pm}(f)$ for complex frequencies in the following manner. Writing $f=\tilde{k}\Omega$ with $\tilde{k} = \tilde{k}_{R} + i \tilde{k}_{I}$, the exponential factor of the integrand can be expanded as
\begin{align}
    \label{eq:Em-complex-expand}
    E_{m}^{\pm}(\tilde{k}\Omega) &= \int_{-\pi}^{\pi} \frac{d\ell}{n} \sum_{q=0}^{\infty} \frac{(-1)^{q}}{q!} \left(\tilde{k}_{I} \tau\right)^{q} e^{\pm i m V} e^{i\tilde{k}_{R} \tau}
\end{align}
where we've defined $\tau = 2\pi \Omega t$. We then use the fact that
\begin{equation}
    \left(\frac{\partial}{\partial \tilde{k}_{R}} \right)^{q} e^{i \tilde{k}_{R} \tau} = (i \tau)^{q} e^{i \tilde{k}_{R} \tau}
\end{equation}
to re-write Eq.~\eqref{eq:Em-complex-expand} as
\begin{equation}
    \label{eq:Em-complex}
    E_{m}^{\pm}(\tilde{k} \Omega) = \sum_{q=0}^{\infty} \frac{(i\tilde{k}_{I})^{q}}{q!} \left(\frac{\partial}{\partial \tilde{k}_{R}} \right)^{q} E_{m}^{\pm}(\tilde{k}_{R} \Omega)\,.
\end{equation}
Since the exponential factor has been eliminated from the integral, one can now apply the approximations of Sec.~\ref{sec:cat} to $E_{m}^{\pm}(\tilde{k}_{R}\Omega)$ in order to obtain the complex valued $E_{m}^{\pm}(\tilde{k}\Omega)$, provided $\tilde{k}_{I} \ll 1$. Eq.~\eqref{eq:Em-complex} combined with Eq.~\eqref{eq:hansen-to-Em} provides us with an analytic approximation of the Hansen coefficients $X^{-3,-m}_{k_{\pm}}$.

\begin{figure*}[tbh!]
    \centering
    \includegraphics[width=\textwidth, trim={3cm 1cm 3cm 2cm}, clip]{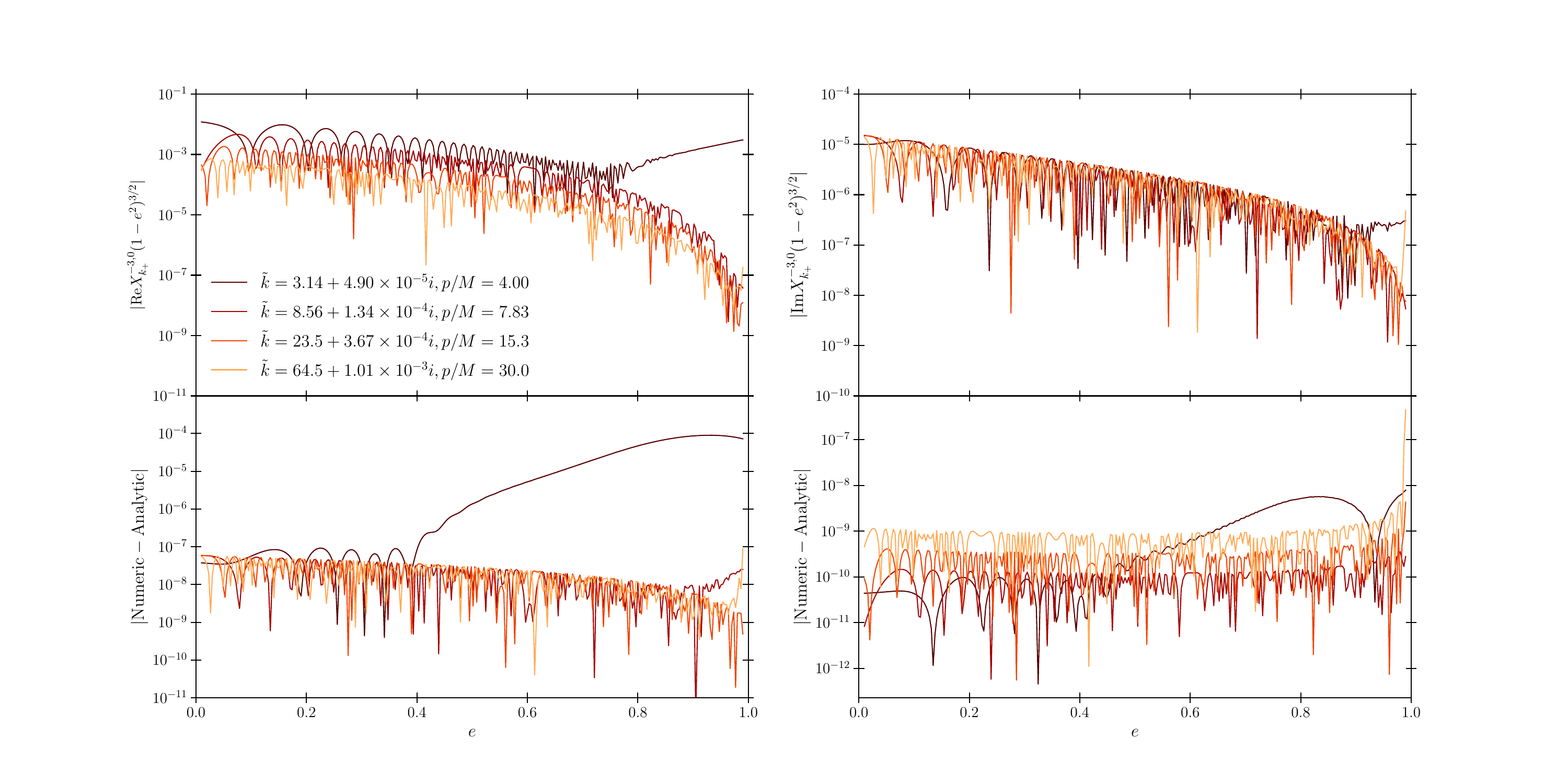}
    \caption{Top: Absolute value of the real (left) and imaginary (right) parts of the Hansen coefficient $X^{-3,0}_{k_{+}}$ as a function of eccentricity $e$. Each color corresponds to a different value of $p/M$, which changes the value of $\tilde{k}$. The f-mode frequency and damping time are the same for each case, specifically $(\omega,\tau) = (6.9 {\rm kHz}, 9.3 {\rm sec})$, corresponding to a $M_{\star} = 1.4M_{\odot}$ NS with radius $R_{\star}=10$ km. Bottom: Difference between the numerical Hansen coefficient and the analytic approximations in Eq.~\eqref{eq:hansen-to-Em} \& \eqref{eq:Em-complex}.}
    \label{fig:hansen_0}
\end{figure*}

\begin{figure*}[tbh!]
    \centering
    \includegraphics[width=\textwidth, trim={3cm 1cm 3cm 2cm}, clip]{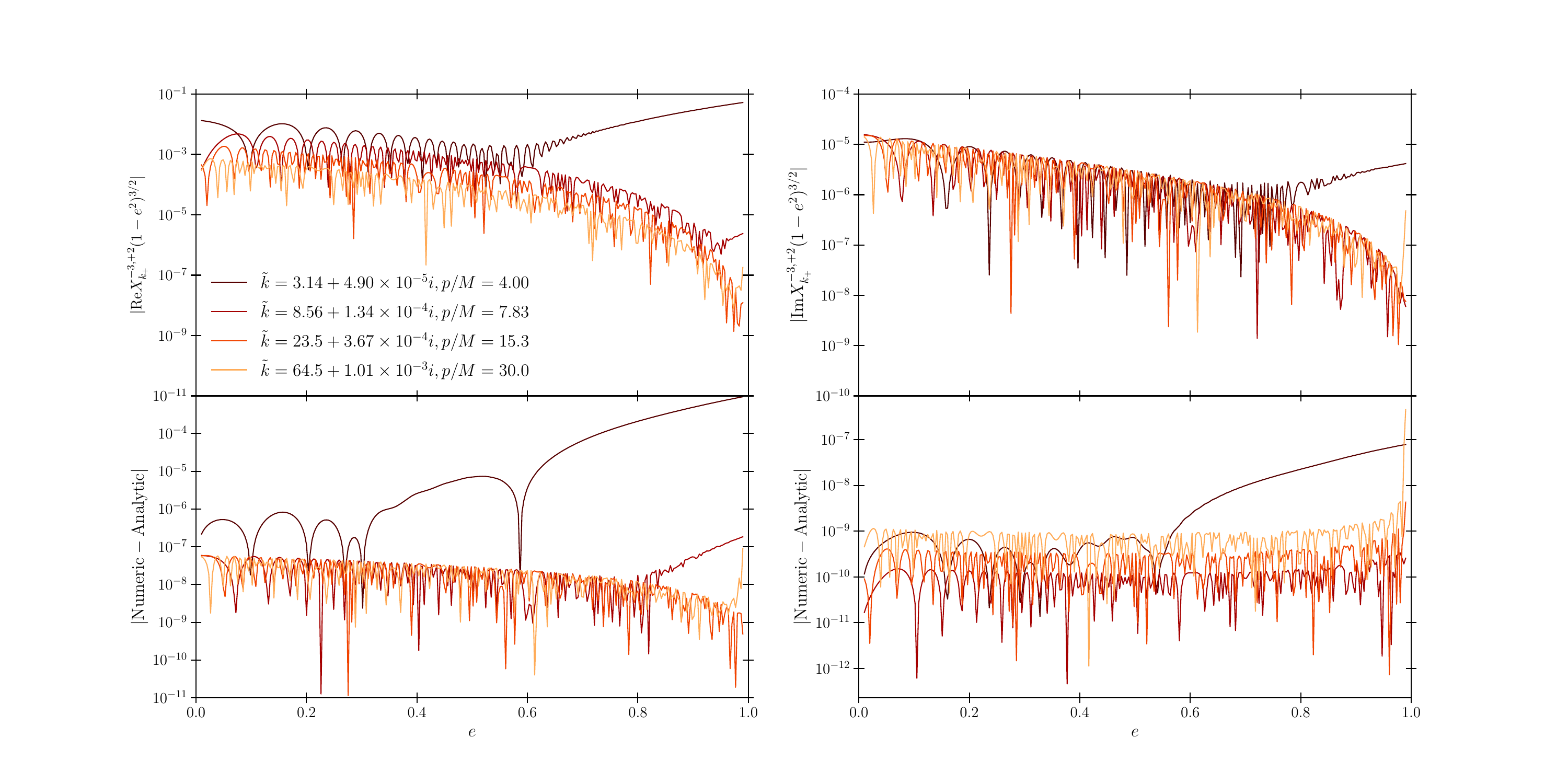}
    \caption{The same as Fig.~\ref{fig:hansen_0}, but for $X^{-3,+2}_{k_{+}}$. }
    \label{fig:hansen_p2}
\end{figure*}

\begin{figure*}[tbh!]
    \centering
    \includegraphics[width=\textwidth, trim={3cm 1cm 3cm 2cm}, clip]{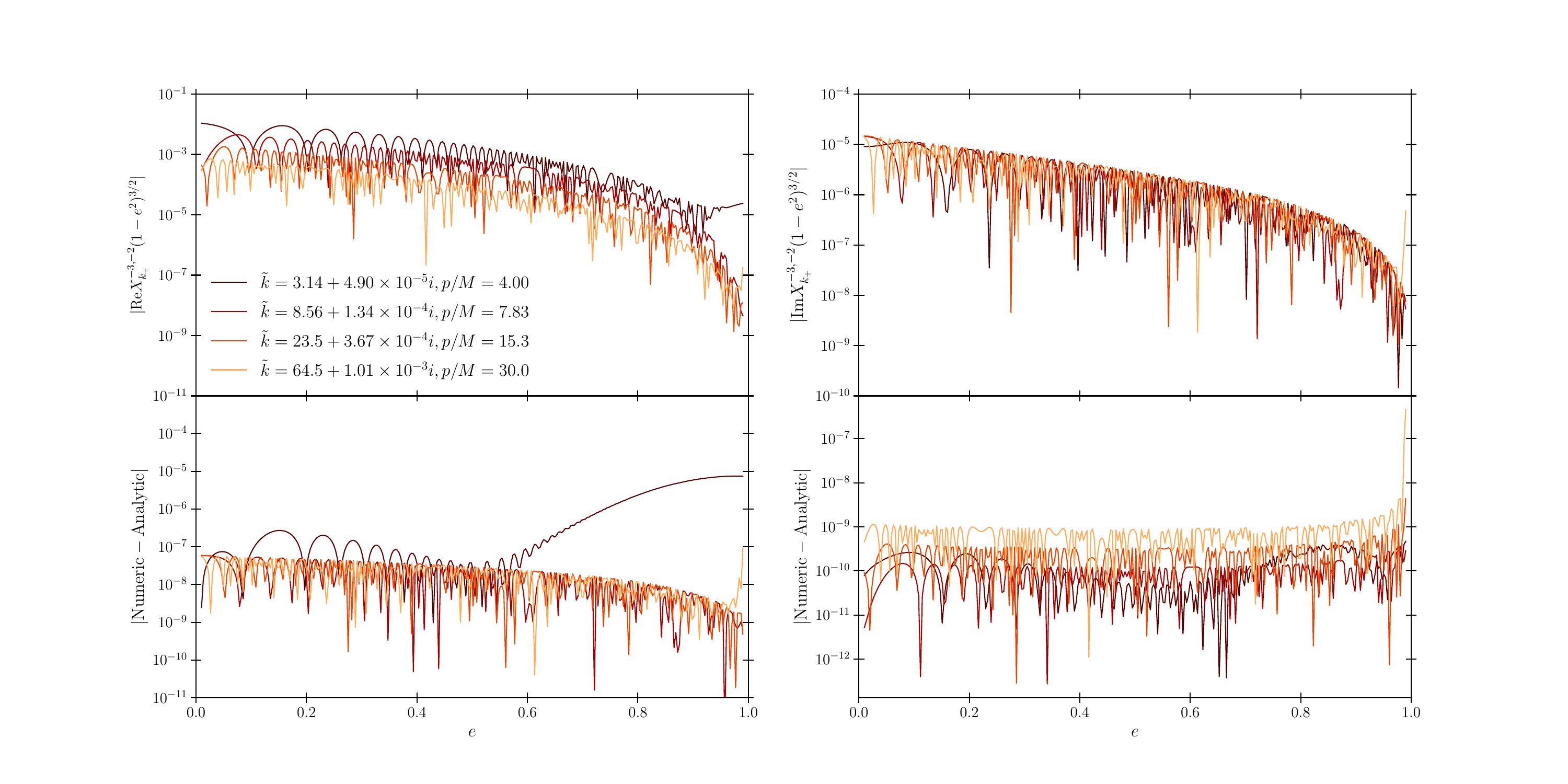}
    \caption{The same as Fig.~\ref{fig:hansen_0}, but for $X^{-3,-2}_{k_{+}}$.}
    \label{fig:hansen_m2}
\end{figure*}

In Figs.~\ref{fig:hansen_0}-\ref{fig:hansen_m2}, we provide a comparison of the analytic approximation of the complex Hansen coefficients to a numerical computation of Eq.~\eqref{eq:hansen-def} with $k=k_{+}$. The results for $k_{-}$ are comparable, and can be found by taking $\tilde{k}_{I} \rightarrow -\tilde{k}_{I}$ in Eq.~\eqref{eq:Em-complex}. To choose proper values of $\tilde{k}$, we decompose this into
\begin{align}
    \tilde{k}_{R} &= {\rm Re}[\tilde{k}] = \frac{\sqrt{\omega^{2} - \gamma^{2}}}{\Omega}\,,
    \\
    \tilde{k}_{I} &= {\rm Im}[\tilde{k}] = \frac{\gamma}{\Omega}\,,
\end{align}
and use $\gamma = 1/\tau$, with $\tau$ the damping timescale. The quantities $(\omega, \tau)$ can be computed using ``universal", or approximately EOS independent, relations found for example in~\cite{Chirenti:2015dda}. Specifically, we use Eqs.~(3) \& (5) therein for $\omega$ and $\tau$, respectively. The quantities $\tilde{k}_{R,I}$ then only depend on the mass $M_{\star}$ and radius $R_{\star}$ of the NS, and $(p,M)$\footnote{Note that~\cite{Chirenti:2015dda} uses $M$ as the mass of the star, which should not be confused with the total binary mass used here.}. For the analysis in Figs.~\ref{fig:hansen_0}-\ref{fig:hansen_m2}, we choose $M_{\star} = 1.4 M_{\odot}$, $R_{\star}=10 {\rm km}$, and $M = 11.4 M_{\odot}$, corresponding to a BHNS binary. The values of the f-mode frequency and damping time for this choice of mass and radius are $\omega = 6.9$ kHz and $\tau = 9.3$ seconds. We allow $p$ to vary between $[4,30]M$ to vary the values of $\tilde{k}_{R,I}$.

Note that the above values of $(\omega,\tau)$ are not representative of results from more realistic NS EOSs. This is due to our choice of $R_{\star}=10$ km, which typically cannot be achieved with our choice of $M_{\star} = 1.4M_{\odot}$ for realistic EOSs. However, we stress that the goal at this point is to test the accuracy of the approximation of the complex valued Hansen coefficients, not to perform an in depth analysis for realistic NSs. The computation herein addresses the lack of analytic approximations for the Hansen coefficients in the f-mode model of~\cite{Arredondo:2021rdt}, and given in Eq.~\eqref{eq:dyn-tide}. We leave computations of realistic NSs and corrections to orbital dynamics, as well as studies of plausible EOS constraints, to future work. 

Each line in Figs.~\ref{fig:hansen_0}-\ref{fig:hansen_m2} provides the result for a different value of $p/M$ and thus $\tilde{k}$. The values of $p/M$ chosen in these figures are those relevant to GWs sources for ground-based detectors. The non-zero imaginary part of $\tilde{k}$ generally causes the Hansen coefficients to obtain a non-trivial oscillation. The top panels of each figure show the numerical value of the Hansen coefficient, computed via the method described in Sec.~\ref{sec:blo}. We do not plot the analytic expression against these, since the difference is typically less than $10^{-4}$, as can be seen by the bottom panels. In all three figures, the case with the largest error has $\tilde{k} = 3.14 + 4.90\times10^{-5}i$, which corresponds to $p = 4M$. Such small values of the semi-latus rectum $p$ are close to the stability limit for geodesics around Schwarzschild BHs as $e\rightarrow 1$~\cite{hansen_stability}. The reason for the error being largest in this case can be seen from Fig.~\ref{fig:Em-comp}. The error in the matched asymptotic expansion is minimized for $f/\Omega >> 1$. Thus, the Hansen coefficients are better approximated by this method for $|\tilde{k}| >> 1$. However, even for the $p=4M$ case, the errors are at most $\sim10^{-5}-10^{-3}$, depending on the value of $m$.
Thus, the complex Hansen coefficients are well approximated by the analytic expressions in Eq.~\eqref{eq:Em-complex}.

While our investigation of Hansen coefficients in this section has focused on those appearing in the f-modes of neutron stars in eccentric binaries, the methodology used here applies for all Hansen coefficients, provided these quantities can be written as sums of the $E_{m}^{\pm}(f)$ functions. As a result, we expect that the analytic approximations developed herein can be applied to more general problems than the few applications considered herein. This completes our investigation of the applications of the results of Sec.~\ref{sec:cat}.

\section{Discussion}
\label{sec:disc}

In this work, we have used catastrophe theory to analyze a fundamental Fourier integral related eccentric Keplerian orbits, and constructed a closed-form, analytic approximation to its response. The methodology presented herein is sufficiently general to be used in any context where fold catastrophes are present in waveform modeling, and could plausibly be extended in the case when higher order catastrophes appear. This is intriguing since the general procedure for dealing with catastrophes in waveform modeling is currently to perform transformations of the waveform to avoid them, resulting in an infinite summation that can be truncated in practice~\cite{Moore:2016qxz,Moore:2018kvz,Klein:2014bua,Chatziioannou:2017tdw}. 

Alternatively, it has been proposed that catastrophe theory could be used to construct full IMR waveforms for quasi-circular binaries~\cite{Jaramillo:2022mkh}, although no practical study of this model in a data analysis setting exists yet. Indeed, much study still needs to be done to determine the efficacy of the catastrophe theory approach and the models developed from it. In the context of eccentric binaries and this work, the primary focus for such studies is the development of full PN waveforms for eccentric bursts, with the general procedure explained in Sec.~\ref{sec:efb}. The pericenter velocity of binaries possessing high eccentricity can reach a sizable fraction of the speed of light, and thus, the inclusion of higher PN order effects into the EFB model developed herein will be necessary. Further, a wealth of studies~\cite{Samsing:2014:BinarySingle,Samsing:2017xmd,SamsigRamirezRuiz:2017:HighlyEccentric,Samsing:2018isx,Samsing:2018ykz,Samsing:2020:AGN,Martinez:2020,Rodriguez:2018pss,Tagawa:2020jnc,Tagawa:2020qll} have shown that binaries with high eccentricity tpyically don't exist in isolation, and are often perturbed by their formation environment. Including such effects in waveform modeling will open the door to extracting astrophysical information about the environments of BBHs from their GW emission~\cite{Romero-Shaw:2022ppk,Sedda:2020:fingerprints,Mapelli:2020:review,Stevenson:2015bqa,Vitale:2015tea}.

Moving beyond the PN setting, the discussion of extending the catastrophe theory analysis in Sec.~\ref{sec:efb} to higher PN order poses an interesting question. Specifically, do similar catastrophes also exist when eccentric binaries are in the dynamical, strong-field regime? The existence of catastrophes in the Newtonian setting at the frequencies $f_{\rm min/max}$ implies the existence of orbital eccentricity through Eq.~\eqref{eq:fmin-max}. If the same holds true in NR simulations and an analogous $f_{\rm min/max}$ can be measured numerically, then one can estimate the orbital eccentricity of binaries in NR simulations. Currently, there are a variety of techniques that attempt to estimate eccentricity in NR simulations. The one specifically used in the SXS catalog~\cite{Boyle:2019kee} requires fitting to a PN inspired model, linearized in small eccentricity~\cite{Buonanno:2010yk}. It would be intriguing to test for the existence of eccentric catastrophes in NR waveforms, and compare eccentricity estimates from these to other measures.

We plan to address these topics in future work. The prospect of using catastrophe theory to understand GWs and develop waveform models does, however, look promising.

\acknowledgements

N.L. acknowledges financial support, in part, provided under the European Union's H2020 ERC Starting Grant agreement no. DarkGRA-757480, the MIUR PRIN and FARE programmes (GW-NEXT, CUP: B84I20000100001), and from the Amaldi Research Center funded by the MIUR program ``Dipartimento di Eccellenza" (CUP: B81I18001170001). N.L. is also supported, in part, by ERC Starting Grant No.~945155--GWmining, 
Cariplo Foundation Grant No.~2021-0555, MUR PRIN Grant No.~2022-Z9X4XS, 
and the ICSC National Research Centre funded by NextGenerationEU. 

\appendix

\section{Coefficients of the MAE}
\label{app:expand}

The MAE of $E_{m}^{-}(f)$ given in Eq.~\eqref{eq:Em-uae} depends on the two functions $[\alpha(f),\beta_{m}(f)]$, which are computed as powers series of the form in Eq.~\eqref{eq:alpha-beta}, with
\begin{equation}
    \zeta = \frac{1}{2e} \left[1 + e - \sqrt{\frac{2\pi f}{m \Omega}}\right]\,,
\end{equation}
where $f\ge 0$. To obtain the power series, one simply has to series expand Eqs.~\eqref{eq:alpha-matched}-\eqref{eq:beta-matched} in $\zeta \ll 1$. The coefficients $[\alpha_{k}, \beta_{k}]$ are functions of the orbital eccentricity, and take the form
\begin{align}
    \alpha_{k} &= (1+e)^{-k} \sum_{j=0}^{k} \alpha_{k,j} e^{j}
    \\
    \beta_{k} &= 4 \left(\frac{m e}{1+e}\right)^{2/3} (1+e)^{-k+1} \sum_{j=0}^{k} \beta_{k,j} e^{j}
\end{align}
Up to $k=7$, the $\alpha_{k,j}$ coefficients are
\allowdisplaybreaks[4]
\begin{align}
    \alpha_{1,0} &= \frac{4}{15}\,, \qquad \alpha_{1,1} = \frac{48}{15}\,,
    \\
    \alpha_{2,0} &= \frac{68}{525}\,, \qquad \alpha_{2,1} = \frac{142}{525}\,, \qquad \alpha_{2,2} = \frac{544}{175}\,,
    \\
    \alpha_{3,0} &= \frac{32}{375}\,, \qquad \alpha_{3,1} = \frac{228}{875}\,, \qquad \alpha_{3,2} = \frac{2164}{7875}\,, \nn \\
    \alpha_{3,3} &= \frac{32096}{7875}\,,
    \\
    \alpha_{4,0} &= \frac{21424}{336875}\,, \qquad \alpha_{4,1} = \frac{86512}{336875}\,, \qquad \alpha_{4,2} = \frac{24356}{61875}\,, \nn \\
    \alpha_{4,3} &= \frac{850592}{3031875}\,, \qquad \alpha_{4,4} = \frac{6122672}{1010625}\,,
    \\
    \alpha_{5,0} &= \frac{339712}{6703125}\,, \qquad \alpha_{5,1} = \frac{16743136}{65690625}\,, \nn\\
    \alpha_{5,2} &= \frac{14515376}{28153125}\,, \qquad \alpha_{5,3} = \frac{104390224}{197071875}\,, \nn\\
    \alpha_{5,4} &= \frac{56901344}{197071875}\,, \qquad \alpha_{5,5} = \frac{3168735616}{328453125}\,,
    \\
    \alpha_{6,0} &= \frac{871637248}{20692546875}\,, \qquad \alpha_{6,1} = \frac{1750090112}{6897515625}\,, \nn \\
    \alpha_{6,2} &= \frac{2643226928}{4138509375}\,, \qquad \alpha_{6,3} = \frac{3573681952}{4138509375}\,, \nn \\
    \alpha_{6,4} &= \frac{2774609072}{4138509375}\,, \qquad \alpha_{6,5} = \frac{6222893536}{20692546875}\,, \nn \\
    \alpha_{6,6} &= \frac{331908564352}{20692546875}
    \\
    \alpha_{7,0} &= \frac{18808049152}{527442890625}\,, \qquad \alpha_{7,1} = \frac{482671223296}{1933957265625}\,, \nn\\
    \alpha_{7,2} &= \frac{896258319728}{1197211640625}\,, \qquad \alpha_{7,3} = \frac{2091159466064}{1676096296875}\,, \nn\\
    \alpha_{7,4} &= \frac{697473501856}{558698765625}\,, \qquad \alpha_{7,5} = \frac{701029684832}{931164609375}\,, \nn\\
    \alpha_{7,6} &= \frac{244815989168}{931164609375}\,, \qquad \alpha_{7,7} = \frac{8524435430416}{310388203125}\,,
\end{align}
and the $\beta_{k,j}$ coefficients are
\allowdisplaybreaks[4]
\begin{align}
    \beta_{1,0} &= 1\,,
    \\
    \beta_{2,0} &= \frac{1}{15}\,, \qquad \beta_{2,1} = -\frac{1}{5}\,,
    \\
    \beta_{3,0} &= \frac{32}{1575}\,, \qquad \beta_{3,1} = \frac{12}{175}\,, \qquad \beta_{3,2} = -\frac{8}{175}\,,
    \\
    \beta_{4,0} &= \frac{656}{70875}\,, \qquad \beta_{4,1} = \frac{344}{7875}\,, \qquad \beta_{4,2} = \frac{682}{7875}\,, 
    \nn \\
    \beta_{4,3} &= -\frac{148}{7875}\,,
    \\
    \beta_{5,0} &= \frac{419392}{81860625}\,, \qquad \beta_{5,1} = \frac{814832}{27286875}\,,
    \nn \\
    \beta_{5,2} &= \frac{96176}{1299375}\,, \qquad \beta_{5,3} = \frac{885616}{9095625}\,,
    \nn \\
    \beta_{5,4} &= -\frac{29584}{3031875}\,,
    \\
    \beta_{6,0} &= \frac{50846144}{15962821875}\,, \qquad \beta_{6,1} = \frac{7776512}{354729375}\,,
    \nn \\
    \beta_{6,2} &= \frac{23501392}{354729375}\,, \qquad \beta_{6,3} = \frac{8203568}{70945875}\,,
    \nn \\
    \beta_{6,4} &= \frac{4878416}{39414375}\,, \qquad \beta_{6,5} = -\frac{1139792}{197071875}
    \\
    \beta_{7,0} &= \frac{10760863744}{5028288890625}\,, \qquad \beta_{7,1} = \frac{727965824}{42976828125}\,,
    \nn \\
    \beta_{7,2} &= \frac{6698868992}{111739753125}\,, \qquad \beta_{7,3} = \frac{14062619392}{111739753125}\,,
    \nn \\
    \beta_{7,4} &= \frac{80482816}{459834375}\,, \qquad \beta_{7,5} = \frac{10137438208}{62077640625}\,,
    \nn \\
    \beta_{7,6} &= -\frac{232181504}{62077640625}\,.
\end{align}
Note that there is no issue with extending the expansions to higher order in $\zeta$ if one desires. We stop here since the accuracy of these approximations are sufficient for the purposes herein, and the expressions for the higher order terms become increasing complicated. 

\section{Approximation of Scorer's Function ${\rm Gi}(x)$}
\label{app:gi}

Scorer's functions ${\rm Gi}(x)$ and ${\rm Hi}(x)$ are the solutions to the inhomogeneous Airy differential equation,
\begin{equation}
    \frac{d^{2}y}{dx^{2}} - x y = \frac{1}{\pi}\,.
\end{equation}
An in depth discussion of these solution can be found at~\cite{NIST:DLMF}. The asymptotic approximation of $E_{m}^{-}(f)$ in Eq.~\eqref{eq:Em-final} relies on ${\rm Gi}(x)$, which is implemented in \texttt{Python} through the \texttt{mpmath} package. We here provide an analytic approximation to this function that significantly speeds up the numerical evaluation of Eq.~\eqref{eq:Em-final}.

\begin{figure}[hbt!]
    \centering
    \includegraphics[width=\columnwidth]{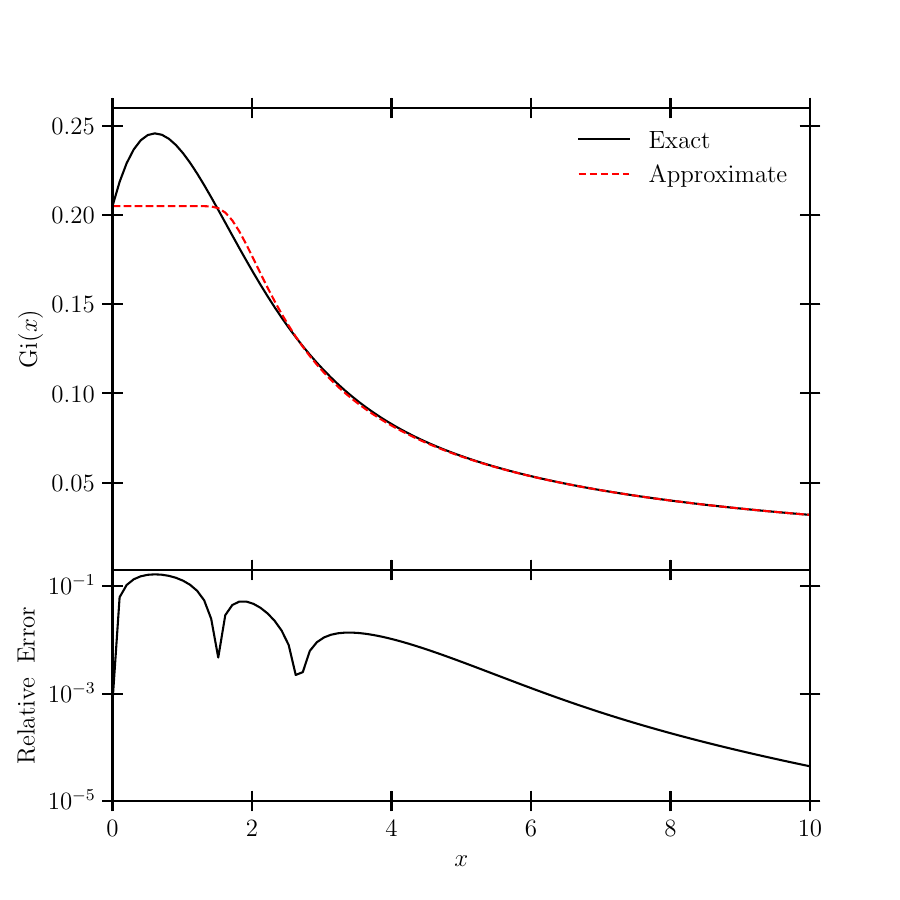}
    \caption{Top: Plot of the exact function ${\rm Gi}(x)$ (solid line) and its approximation in Eq.~\eqref{eq:gi-approx}. Bottom: Relative error between the exact function and the approximation.}
    \label{fig:gi}
\end{figure}

The starting point is the asymptotic expansion of ${\rm Gi}(x)$, specifically
\begin{equation}
    \label{eq:gi-asym}
    {\rm Gi}(x) \sim \frac{1}{\pi x} \sum_{k=0}^{\infty} \frac{(3k)!}{k! (3x^{3})^{k}}\,.
\end{equation}
Note that this expansion holds when $x \rightarrow \infty$, and the function ${\rm Gi}(x)$ is regular at $x=0$. The goal is to obtain a new function that approximates the behavior of ${\rm Gi}(x)$ in the domain $0\le x < \infty$. There are multiple functions that one can construct to fit this type of behavior, an example being
\begin{equation}
    \label{eq:gi-approx}
    [{\rm Gi}(x)]_{\rm approx} = g_{0} \left[1 - e^{-\sum_{k=1}^{\infty} g_{k} x^{-k}}\right]\,.
\end{equation}
By expanding the above expression about $x\rightarrow \infty$, the coefficients $g_{k}$ can be matched to the coefficients of the series in Eq.~\eqref{eq:gi-asym}. Up to $k=7$, the coefficients are
\allowdisplaybreaks[4]
\begin{align}
    g_{0} &= {\rm Gi}(0) =  \frac{1}{3^{7/6}\Gamma(2/3)} = 0.204975542482000\,,
    \\
    g_{1} &= \frac{1}{\pi g_{0}}\,, \qquad g_{2} = \frac{1}{2\pi^{2} g_{0}^{2}}\,, \qquad g_{3} = \frac{1}{3\pi^{3} g_{0}^{3}}\,, 
    \\
    g_{4} &= \frac{1}{4\pi^{4} g_{0}^{4}} + \frac{2}{\pi g_{0}}\,, \qquad g_{5} = \frac{1}{5\pi^{5} g_{0}^{5}} + \frac{2}{\pi^{2}g_{0}^{2}}\,,
    \\
    g_{6} &= \frac{1}{6\pi^{6} g_{0}^{6}} + \frac{2}{\pi^{3}g_{0}^{3}}\,, \qquad g_{7} = \frac{1}{7\pi^{7}g_{0}^{7}} + \frac{2}{\pi^{4} g_{0}^{4}}\,,
\end{align}
which constitutes the approximation we use for ${\rm Gi}(x)$. Fig.~\ref{fig:gi} provides a comparison between the exact function computed using \texttt{mpmath} and the analytic approximation, with the bottom panel showing the relative error. The approxiation in Eq.~\eqref{eq:gi-approx} is roughly one hundred times faster to evaluate than the exact function from \texttt{mpmath}, and does not result in a significant loss of accuracy in the applications studied in Sec.~\ref{sec:apps}.

\section{Asymptotic Expansion of $E_{m}^{+}(f)$}
\label{app:Em_plus}

We here provide an explicit computation of $E_{m}^{+}(f)$ for $f \ge 0$. As is explained in Sec.~\ref{sec:cat}, for negative frequencies $E_{m}^{+}(-f) = \left[E_{m}^{-}(f)\right]^{\dagger}$, and is thus approximated using the methods therein. For positive frequencies, there are no stationary points in the domain of integration, and the saddle point at $V=0$ becomes subdominant since the integrand in Eq.~\eqref{eq:Em-int} becomes highly oscillatory in the region around this point. The integral is actually dominated by the region near $V = \pm \pi$, in contrast to the behavior of $E_{m}^{-}(f)$. To analytically approximate $E_{m}^{+}(f)$, it is useful to re-write the integral as
\begin{equation}
    E_{m}^{+}(f) = 2\int_{0}^{\pi} \frac{d\ell}{n} \cos\left[\Psi_{m}^{+}(t,f)\right]\,,
\end{equation}
with
\begin{equation}
    \Psi_{m}^{+}(t,f) = 2\pi f t + m V(t)\,.
\end{equation}
Due to the lack of stationary points in the domain of integration, this integral can be evaluated in an asymptotic expansion by repeated integration by parts using the same method described in Sec.~\ref{sec:blo}. However, unlike the case of ${\cal{R}}(f)$, one does not need to Taylor expand $\Psi_{m}^{+}$. The integral can be directly evaluated using the fact that
\begin{equation}
    \cos\left[\Psi_{m}^{+}(t,f)\right] = \frac{1}{\dot{\Psi}_{m}^{+}(t,f)} \frac{d}{dt} \sin\left[\Psi_{m}^{+}(t,f)\right]\,.
\end{equation}
To leading order,
\begin{equation}
    \label{eq:Ep-approx}
    E_{m}^{+}(f) \sim \frac{\sin\left(m \pi + \frac{2\pi^{2}f}{n} \right)}{\pi (f + f_{\rm min})} + {\cal{O}}\left[(f+f_{\rm min})^{-4}\right]\,.
\end{equation}
We truncate the expansion at leading order due to the fact that the higher order corrections constitute a divergent series, and including them doesn't necessarily improve the accuracy of the approximation.

\bibliographystyle{apsrev4-1}
\bibliography{refs}
\end{document}